# Choice Architecture, Privacy Valuations, and Selection Bias in Consumer Data

Tesary Lin[*]     Avner Strulov-Shlain[†]

August 21, 2023

**Abstract**

We study how choice architecture that companies deploy during data collection influences consumers' privacy valuations. Further, we explore how this influence affects the quality of data collected, including both volume and representativeness. To this end, we run a large-scale choice experiment to elicit consumers' valuation for their Facebook data while randomizing two common choice frames: default and price anchor. An opt-out default decreases valuations by 14-22% compared to opt-in, while a $0–50 price anchor decreases valuations by 37-53% compared to a $50–100 anchor. Moreover, in some consumer segments, the susceptibility to frame influence negatively correlates with consumers' average valuation. We find that conventional frame optimization practices that maximize the volume of data collected can have opposite effects on its representativeness. A bias-exacerbating effect emerges when consumers' privacy valuations and frame effects are negatively correlated. On the other hand, a volume-maximizing frame may also mitigate the bias by getting a high percentage of consumers into the sample data, thereby improving its coverage. We demonstrate the magnitude of the volume-bias trade-off in our data and argue that it should be a decision-making factor in choice architecture design.

*Keywords:* privacy, choice architecture, market for data, selection bias, experiment

[*]Boston University Questrom School of Business; tesary@bu.edu

[†]University of Chicago Booth School of Business; avner.strulov-shlain@chicagobooth.edu. The authors thank Guy Aridor, Dan Bartels, Josh Dean, Berkeley Dietvorst, Kwabena Donkor, Andrey Fradkin, Sam Goldberg, Ali Goli, Avi Goldfarb, Michael Grubb, Tanjim Hossain, Yufeng Huang, Alex Imas, Jihye Jeon, Garrett Johnson, Yucheng Liang, Nina Mazar, Sanjog Misra, Ilya Morozov, Olivia Natan, Omid Rafieian, Heather Sarsons, Ananya Sen, Tim Simcoe, K. Sudhir, Oleg Urminsky, Giorgos Zervas, and Jinglong Zhao for their helpful comments. We thank Christy Kang, Paulina Koenig, and Kaushal Addanki for their excellent research assistance. This research is funded by the Becker Friedman Institute at the University of Chicago and the Willard Graham Research Fund at Chicago Booth. It was approved by the Institutional Review Boards at the University of Chicago (IRB21-1376) and Boston University (IRB-6239X).

# 1 Introduction

How companies should collect and use consumers' personal data is at the center of recent policy debates. Companies often deploy some form of "choice architecture" (Thaler & Sunstein 2008) when collecting consumer data, which are choice environments designed to nudge consumers towards sharing more data, all else equal. [1] For instance, after the GDPR took effect, websites have been using a combination of default settings, salient options, and obstructions to nudge their users toward sharing cookie identifiers (Matte et al. 2020, Nouwens et al. 2020). Several major consent management platforms, such as OneTrust and Usercentrics, offer products and resources to help websites design user interfaces that will maximize opt-ins.[2]

Existing choice architecture optimization practices emphasize maximizing the volume of data collection as their end goal. What is often neglected is a second dimension of data quality—the representativeness of the data collected. Biased input data often leads to biased insights and decision-making. For instance, Cao et al. (2021) show that gender bias in Product Hunt's product votes leads to severe bias in the predicted appeal of new products, causing female-focused products to experience 45% less growth than male-focused counterparts. As another example, Bradley et al. (2021) show that using large but unrepresentative samples leads to overestimates of COVID-19 vaccine uptake by 14-17 percentage points. Bias in shared consumer data can often ensue when different consumer groups have different valuations of their data (Lin 2022). However, an underappreciated aspect is how choice architecture affects the representativeness of data shared when different types of consumers respond to choice architecture to different degrees.

In this paper, we ask the following questions: How do choice frames influence consumers' privacy valuations, and what is the heterogeneity of the choice frame effects? How do the choice frames change the composition of consumers sharing their data, beyond its influence on the quantity of data shared? The economic returns of data to firms depend on both its quantity and its representativeness. Therefore, to assess how choice architecture affects the quality of data collected and the efficiency of data collection, we must account for its effect on both the volume and the composition of data shared.

To answer these questions, we recruited 5,028 Facebook users, and elicited their willingness-to-accept (WTA) to sell their data while randomizing the choice frames they faced. Within participants, we asked their valuation for sharing the following variables with researchers and advertisers: *about me* (their information on the profile page), *posts*, *likes*, *friends and followers*, and *responses to our survey*. For each variable, we elicited incentive-compatible WTA using a multiple price list (Kahneman et al. 1990, Andersen et al. 2006), followed by a free-text entry. Across participants, we randomized the choice default and the price anchor. *Default* varies between opt-in, opt-out, and active choice. *Price anchor* is the range of prices in the multiple price list, which is either $0-$50 (low) or $50-$100 (high). We choose default and price anchor because they are common or likely to be deployed if companies can buy data from consumers.[3] We also collect consumer characteristics to explore heterogeneity in their privacy valuations and responses to choice frames. These variables include demographics, social media and Internet usage, and their belief about what data are already available to Facebook and the public.

---
[1] We use "choice architecture", "choice frames", and "frames" interchangeably throughout the paper.
[2] https://www.onetrust.com/blog/onetrust-launches-consent-rate-optimization-to-maximize-opt-ins/; https://usercentrics.com/resources/opt-in-optimization/

Consumers' valuations for data are substantially different across both individuals and data. Across individuals, valuations for the same data range from $0 all the way up to infinity, with 20% having their valuations above $100, the maximal value that we offer in the multiple price list. When we top-code the values at $100, the mean valuation across data is $63.9. The difference in median WTA between the most valuable (friends and followers) and the second most valuable data (posts) is $8.9, while the difference between the most and the least valuable data (survey responses) is $28.4.

We also find a significant influence of choice frames on consumers' valuations. Consumers decrease their valuations by 37.4%-52.6% in the *low* versus *high* price anchor condition, depending on how we topcode the data. Compared to an *opt-out* default, *active choice* increases the valuation of data by 5.8%-11.8%, and *opt-in* increases the valuation by 13.6%-21.1%. On average, the difference in valuation is $16.2 lower due to the *low* price anchor, and is $2.2 and $5.1 higher in the *active* and *opt-in* defaults when we use the $100 topcode. Although the qualitative effect of defaults is uniform, the price anchor distorts data valuations in complex ways. Valuations bunch towards the endpoints of a price range. At the same time, the proportion of consumers who report extremely high values (e.g., "I do not want to share my data at any price") increases from 15.5% to 21.2% as a result of the low price anchor. This pattern suggests that a low anchor can cause "backlash" among some consumers while decreasing the valuation of others.

To explore how choice frames affect the composition of the shared data, we estimate causal forest models (Athey et al. 2019) to see what consumer attributes correlate with the heterogeneity in data values and frame effects. The causal forest models allow us to nonparametrically and efficiently characterize the joint distribution of privacy valuations and frame effects. We find that consumers' valuations of their data and their responses to choice frames are negatively correlated overall. Such a negative correlation is stronger across specific consumer segments. Consumers who value their data less across frames are overall younger, poorer, less educated, and more likely to click on ads while on Facebook. Interestingly, these attributes also predict larger frame effects.

The negative correlation between privacy valuation and choice frame effects creates a potential trade-off between volume-maximizing and bias-minimizing objectives during data collection. When a firm can choose the choice frame while buying data, it often wants to adopt a frame that maximizes the volume of data collected for a given price. Such a volume-maximizing frame has the potential to exacerbate the statistical bias in the data collected by the firm: Consumer groups who already value their data less absent frame effects now give up their data even more willingly due to the frame. In this case, the collected data may oversample this group even more, while possibly undersampling other consumer groups if the firm also sets a lower price for data due to the supply expansion. A biased dataset compromises the quality of data-driven analytics and inferences. In fact, a large yet biased dataset can lead to overly precise yet wrong estimates, misguiding business and policy decisions (Bradley et al. 2021). As such, the potential bias induced by a volume-maximizing choice frame can counteract the benefits of collecting more data and decrease the value of shared data as a result.

---

[3] Our method for eliciting privacy values may differ from typical corporate data request practices. Nevertheless, it is a practical demonstration of how the frame effect heterogeneity influences the quality of shared data.

We perform counterfactual analysis to explore the volume-bias trade-off. We start by comparing the quality of data collected under the volume-maximizing (hereafter *vol-max*) frame with a frame-average benchmark, where we average the valuations across different frames in our experiment, approximating the outcome when the firm chooses among the available frames at random. We show that the vol-max frame can have countervailing effects on the bias in collected data compared to the benchmark. With a negative correlation between privacy valuation and choice frame responsiveness, the vol-max frame may exacerbate the bias in sample data. In particular, holding fixed the sample size target, the vol-max frame tends to collect data with more bias compared to the benchmark frame. On the other hand, increasing the volume of data can be useful for mitigating the bias: For example, imagine that at a certain point all the consumers in the shared data are low-type consumers. As the volume of sample data continues to grow, the consumers newly added to the sample data will all be high type and thus alleviating the bias. Since the vol-max frame gathers more data at any given price point compared to the benchmark frame, the impact of the vol-max frame on bias in sample data is ambiguous if the equilibrium price for data does not move much. We show that the bias-mitigating effect tends to dominate when the percentage of consumers sharing their data is large and when the firm's demand for data is elastic.

We then demonstrate this trade-off from another angle by comparing the vol-max and bias-minimizing (hereafter *bias-min*) frames. For each given price, we identify the choice architecture that maximizes the volume and the one that minimizes the bias, and characterizes the sample data collected under each frame. On average, moving from the volume-maximizing choice architecture to the bias-minimizing one leads to a 46.5% (sd = 25.2%) reduction in bias across covariates, and a 7.7% (sd = 4.9%) reduction in volume. These numbers translate to an elasticity of volume to bias-reduction of 0.17. At the same time, the median volume-to-price elasticity is 0.70. In other words, to reduce bias by 10%, the firm can either choose an alternative frame that reduces data volume by 1.7% under the same price for data, or raise price by 2.4% in order to not sacrifice volume.

The magnitude of the trade-off shifts substantially once we allow the firm to personalize its frame assignment. With personalized frames, shifting from vol-max to bias-min leads to a 71.4% reduction in bias accompanied by a 9.0% reduction in volume; the corresponding volume-to-bias elasticity drops from 0.16 to 0.13. This result is driven by the fact that in our setting, the frame that shifts consumer valuation down the most is mostly the same (low anchor + opt-out), while personalization brings a more substantial gain in bias reduction.

Our paper contributes to the existing literature on several fronts. Our main contribution lies in showing how the heterogeneous effects of choice architecture create selection issues in the data market. We build on the literature on measuring consumers' valuation of privacy using a revealed preference approach (Goldfarb & Tucker 2012, Athey et al. 2017, Kummer & Schulte 2019). Although existing literature has documented the presence of privacy preferences, empirical studies quantifying the value of data to consumers are nascent (Acquisti et al. 2013, Lin 2022, Tang 2019, Collis et al. 2020). One potential reason is that privacy preferences are context-specific (Martin & Nissenbaum 2016) and hard to measure. Lin (2022) highlights consumers' economic reasoning in different data usage scenarios (the *instrumental preference*) as a contributor to the context effect. In comparison, we focus on the behavioral contributor to the context effect—the choice architecture, and explore how it affects the composition of consumers who share data.

A wealth of literature has examined how choice architecture affects privacy choices (Johnson et al. 2002, Acquisti et al. 2012, 2013, Adjerid et al. 2019, Kormylo & Adjerid 2021, D'Assergio et al. 2022, Tomaino et al. 2022). Our innovation lies in relating the frame effects to how consumers self-select into sharing data and the quality of data available to firms. As such, our focus connects two disjoint threads of literature: behavioral biases in privacy valuations, and the efficiency of data markets (Arrieta-Ibarra et al. 2018, Acemoglu et al. 2022, Bergemann et al. 2022, Ichihashi 2021, Markovich & Yehezkel 2021). We also contribute to the empirical literature on behavioral industrial organization that takes consumers' biases and fallibility into firms' decision-making processes (Peltzman 1981, Rao & Wang 2017, Strulov-Shlain 2023, Miller et al. 2022).

The heterogeneity of choice architecture effects has broader relevance than in the market for data (Mrkva et al. 2021). Whenever there are distributional concerns, such as in government programs (Linos 2018, Misra 2023) or marketing campaigns (Dubé & Misra 2023), the heterogeneous effects of choice architecture and how it relates to the heterogeneity of baseline propensities will matter: A well-chosen choice architecture can help balance reach and distributional goals.

Our project also contributes to the literature on the value of consumer data to firms (Rossi et al. 1996, Trusov et al. 2016, Miller & Skiera 2017, Bajari et al. 2019, Aridor et al. 2021, Rafieian & Yoganarasimhan 2021, Sun et al. 2021, Wernerfelt et al. 2022, Lei et al. 2023, Peukert et al. forthcoming) by highlighting the importance of considering data composition when evaluating its value. Iansiti (2021) theorizes that the marginal value of data is initially very high as firms try to overcome the model's "cold-start" phase. However, diminishing marginal returns quickly set in, after which a firm's competitive advantage of data ownership mainly comes from having unique data points. In our framework, this theory implies that a firm may initially place more weight on increasing data volume, but will eventually shift to getting a more representative dataset with broader coverage to gain a competitive edge.

Lastly, our work is also connected to the recent research on how bias in input data creates biased algorithms (Cao et al. 2021), biased estimates of public opinions and behaviors (Bradley et al. 2021), degraded performance in business analytics (Lin 2022, Neumann et al. 2022), and other market outcomes (Johnson et al. 2020). Although the sources of input data bias vary, individual differences in privacy concerns is often one of them. Our work explores this angle further by examining how a choice architecture chosen by the firm may exacerbate or alleviate this bias.

The rest of the paper is organized as follows. Section 2 uses a conceptual model to illustrate why accounting for the joint heterogeneity of privacy valuation and frame effect is crucial for understanding how frames affect bias in the sample data. Section 3 illustrates the design of our experiment. Section 4 describes our data and reduced-form evidence, followed by heterogeneity analysis for the privacy valuations and choice frame effects. Section 5 introduces our counterfactuals to show when a volume-maximizing frame may create the bias-volume trade-off, and Section 6 concludes.

## 2 Conceptual Framework

In this section, we lay out the framework that underpins our experimental design and subsequent analysis. We start by distinguishing two types of data valuation. The first is a hypothetical frame-neutral valuation $v_0$, which is the would-be valuation if all choice frames were absent. This valuation reflects consumers' best guess about the true value of their privacy. We note that this true value is not observed by the researcher or the consumer. The observed valuation is what we call the expressed valuation $\widetilde{v}$, which is a function of the frame-neutral valuation but is influenced by choice architecture, $\theta$:

$$\widetilde{v} = f(v_0; \theta). \tag{1}$$

In other words, the choice architecture creates a gap between the expressed valuation and the consumers' prior judgment of what the value of their data is.

We situate the two valuations in the context of a data market, where firms directly interact with consumers to get their consent for sharing data. In such a market, consumers form their supply for data based on their privacy valuations; the firm has a demand for data, offers a price for buying data, and can choose the choice frame to influence the supply from consumers.[4] Although $\widetilde{v}$ may be larger or smaller than $v_0$, the firm often chooses a frame that delivers the lowest $\widetilde{v}$ to maximize its gain per dollar offered. In essence, the firm-chosen frame pushes the consumers' data supply curve toward the right (see Figure 1a). As a result, the equilibrium quantity of data traded is larger and the equilibrium price is lower compared to the frame-neutral equilibrium.

---

[4]The price for consumer data can either be explicit or implicit. For example, a form of implicit price can take the form of access to services or personalization benefits upon data sharing.

Figure 1: Distortion of Data Market Due to Choice Frames: An Illustration

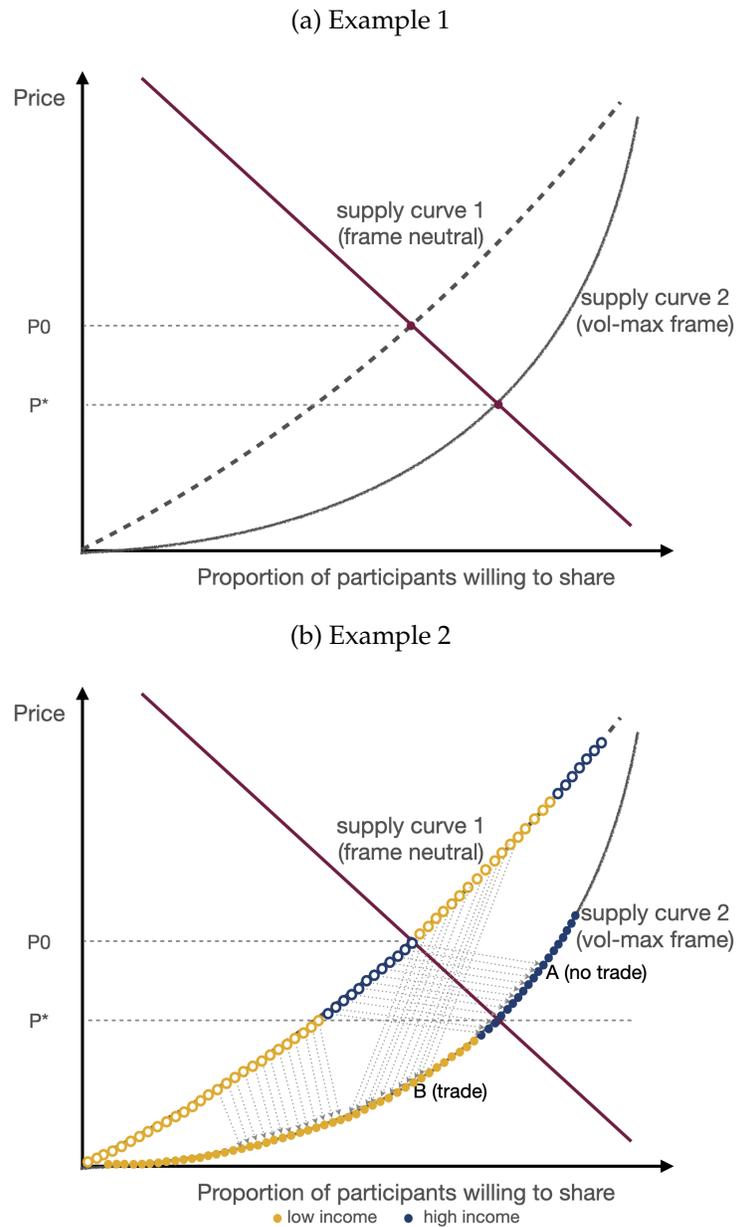

(a) Example 1

(b) Example 2

So far, we have shown how choice frames can induce behavioral distortions and change data collection by treating data as a standard commodity. However, consumer data is not a standard commodity. One unique feature of data is that its value to firms depends on not only its volume but also its representativeness. When evaluating the impact of choice architecture, the firm should care about not just its impact on the volume of data shared, but also which consumers are more likely to trade and how that affects the bias in the data sample.

How does the consideration of sample bias affect our evaluation of frame effects? Consider the following example shown in Figure 1b. Absent the choice frame effects, low-income consumers value their privacy less compared with wealthier ones; however, their valuations are also more susceptible to the influence of choice architecture. If a frame-neutral choice architecture exists, then for any given price the firm would have under-sampled wealthier consumers while over-sampling poorer ones. With a choice frame that pushes the data supply downward, such selection bias becomes more severe because poorer consumers adjust their valuations downward even further compared to the wealthier ones. A trade-off emerges: The volume-maximizing choice frame would have helped the firm in a regular commodity market, but can end up harming the firm by inducing more bias in the collected data.

In our illustrative example above, deploying a volume-maximizing frame ends up exacerbating bias in the sample data due to the negative correlation between consumers' privacy valuations and their frame responses. However, we note that in empirical settings, this need not always happen. For instance, suppose all consumers who share their data in the neutral frame belong to the low-income segments. In this case, deploying the volume-maximizing frame will never exacerbate the bias (it was maximally biased to begin with), and may even alleviate the bias by including more consumers in the sample. In other words, the volume-maximizing frame can also have a bias-mitigation effect through supply expansion.

To examine the potential bias-volume trade-off in practice, we use an experiment to randomize choice frames between participants, while collecting consumer characteristics to explore the heterogeneity in their privacy valuations as well as frame effects. This approach allows us to construct the empirical analog of Figure 1a, and unpack how the trade-off present in frame choices depends on demand and supply conditions in a data market.

## 3 Experiment

The main component of our experiment is a multiple price list (MPL) that elicits consumers' valuation of their Facebook data in an incentive-compatible fashion. To test the effects of choice frames on reported valuation, we randomize the default choice and the range of price list between participants when implementing the MPL. We also use baseline and endline questions to measure consumer characteristics, their internet and social media usages, and their beliefs about which of their data are already available. Figure 2 summarizes the flow of our experiment. Below, we start with the participant recruiting procedure, then introduce the value elicitation components, the choice frame treatments, and end with the design of survey questions that measure consumer characteristics.

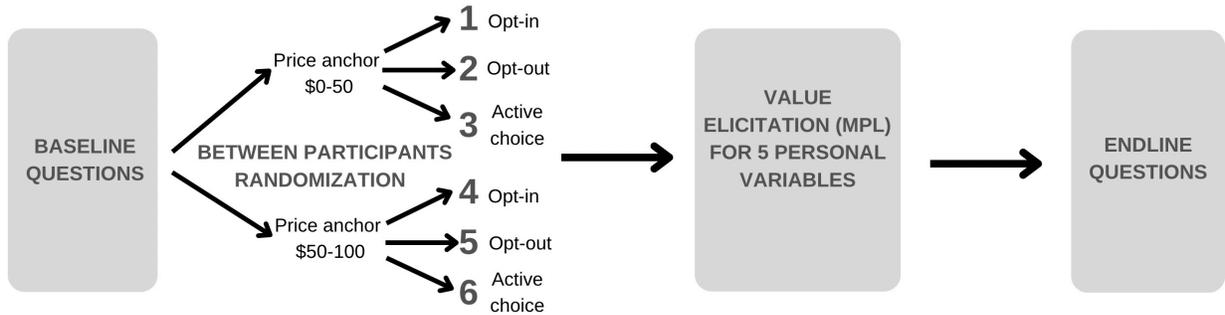

Figure 2: Experiment Overview

## 3.1 Participant Sources

We recruit our participants from two sources: Facebook Ads and Prolific. Using these sources confers two advantages. First, both sources allow us to screen participants based on the availability of their Facebook accounts.[5] Including only participants who have an active Facebook account ensures that their data sharing decision reflects only their privacy preferences and not the availability of their data. Second, including participants from both sources facilitates external validity analysis. Existing studies that measure privacy preference often use survey panels or student populations, who may have lower valuations for privacy or respond to choice frames differently compared with the population. Having both the Facebook and survey panel (Prolific) participants allows us to examine this possibility. We restrict our participants to English speakers living in the US, with an age between 18 and 64.

To minimize selection into the experiment, we do not disclose the specific research topic in the recruiting ad (for the Facebook participants; see Figure A.1) or the study invitation email (for Prolific). A user who clicks on the survey invitation link from the recruiting ad or email is directed to our study introduction page. The introduction explains that we are university researchers who want to understand the public's social media usage and perceptions,[6] then ask for their consent to enter the study.

## 3.2 The Multiple Price List and Choice Frame Treatments

We measure participants' valuation of their Facebook data using a multiple price list (MPL). MPL resembles Becker-DeGroot-Marschack (Becker et al. 1964) in its use of a lottery to ensure incentive compatibility. Its advantage is simplicity: Since MPL uses simple take-it-or-leave-it offers repeated at different price points instead of a second-price auction, it is easier for participants to understand the procedure and why telling the truth is optimal. Standard MPLs give us only the interval that includes a consumer's valuation of their data (see Figure A.3. To get more granular numbers, we follow each MPL with an open-text question asking participants their exact valuation of the data. If a participant chooses not to share data in all the MPL questions displayed, we also give them the option to indicate "I do not want to share my data at any price" in the open-text prompt (see Figure A.4). Suppose a consumer's response is chosen by the lottery. In this case, the data-money exchange occurs if the number in their free-text entry is lower than or equal to the offer price randomly generated by the computer. This mechanism guarantees that reporting the true value is a dominant strategy.

---

[5] When recruiting the Facebook participants, we restrict our ad placement to positions only viewable by logged-in Facebook users, which excludes Facebook Audience Network and Instagram. For the Prolific participants, we use the internal selection tool provided by the platform to enlist only Facebook users.

[6] This information includes their data sharing behavior and privacy attitudes on social media, but to minimize selection bias, we did not explicitly mention that on the introduction page.

After answering the baseline questions, participants see a message asking for their valuation to share data with advertisers. An example of the MPL procedure follows, showing them how their choices and the random price generated by the computer co-determine whether the data exchange will occur:

> Your answers to the survey questions, and other information, can help us understand browsing behavior better. It can also help companies and advertisers provide more products that they think you like, and show you fewer products that you are less likely to buy. **Would you be willing to share more data with us and advertisers?**
>
> **If you will, we will pay you a fair price.** When you started the survey, the computer already randomly selected if you will be asked to provide data at the end of the survey, and also randomly chose a price we will pay for it. If you are selected to participate, your payment and data shared with advertisers will base on what you choose. So you should answer carefully!
>
> For example: suppose you choose to share some data for a price of $Y. If the computer chose a price lower than $Y, you will not be asked to share the data and will not be paid. If the computer chose a price larger than $Y, you will be asked to download a copy of your data from Facebook and send them to us; you will then get the price the computer has chosen.

Participants then go through a practice question, where we ask them to imagine selling a gift card worth $14.5 by responding to a multiple price list. If a participant gives a value different from $14.5, we display an error prompt, asking them to think again and showing how a truthful response is optimal (see Figure A.2). Since the price anchor treatments do not span the entire range, we also train participants to report values outside their allotted treatment and to get familiar with the exact elicitation interface. For example, a participant in the low price range treatment (seeing an MPL with a price range of $0-$50) will be asked to report the value of selling a hypothetical $60 gift card. To do so, they need to choose "No" for all prices in the MPL; they should then type in $60 in the free-text question. Similarly, the high price range participants need to agree to share with all prices greater or equal to $50 and then report $40 as the free-text response.

After the practice round, we show participants the actual MPL questions to get their valuation for different types of data. We inform participants that our offer price is randomly drawn between $0-$95 and does not depend on their response. Each participant receives five rounds of MPLs in random order, one for each personal variable. The following list shows the five variables and the definition we show to participants:

- *"About me" page: your Facebook information page;*
- *Posts: your Facebook posts and feed history;*
- *Likes: the posts and pages you liked on Facebook;*
- *Friends and followers: the people you befriended and followed, and people who followed you;*
- *Survey answers: the answers you gave earlier on your browsing behavior and demographics.*

We independently vary the following choice frames across participants. The first frame is the **default** choice in the multiple price list. A pre-selected "yes" for the question "will you share your posts for $50" is an *opt-out* default, while a pre-selected "no" is *opt-in*. For *active choice*, neither option is pre-selected; participants will have to click on one of them to answer the questions and proceed to the next screen (see Figure A.3. The second frame is the **price range** offered in MPL: the *low price* condition has prices between $0 and $50, while the *high price* condition ranges from $50 to $100. These choice frames are always the same within participant.

Apart from the choice frames, we also randomize the time range of behavioral data (*posts* and *likes*) between participants, which varies between *1 month*, *1 year*, and *since joining Facebook*. In doing so, we vary the value of data in a direction known to us. The goal is to see if consumers respond to the scope of data requested when valuing their personal data.

We inform participants that their responses will be chosen to implement the data-price exchange with positive probability. All participants receive a flat participation fee immediately after finishing the survey: We ensure they receive the participation payment in time to build trust that we will also pay them if they are selected by the lottery and send us their data. For participants chosen by the lottery and with valuations lower than our randomized price, we sent them a step-by-step guide to download the Facebook variable. They receive the payment within 24 hours after sending their data to us.

### 3.3 Baseline and Endline Questions

The survey includes questions that capture consumer characteristics, their internet and social media usage, as well as their beliefs about personal data availability. In general, we put questions related to data sharing in the endline survey, so that participants are not primed to consider privacy before the MPL questions; otherwise, we include the questions in the baseline survey.

The baseline questions measure participants' social media consumption behavior and their demographics. We ask participants about their time spent on Facebook and online, when they started using Facebook, and their engagements with merchants and ads on Facebook. For demographics, we record their age, gender, ethnicity, income, and education.

The endline questions include measures of information-seeking behavior and participants' beliefs about data availability. To measure information seeking, we include the following question: *"Did you look up additional information when answering the questions that ask how much you value your Facebook data?"* If they answer "yes", we ask what kind of information they looked up. To measure their belief about data already available to various parties in the market, we ask the following questions: (a) *Which of your personal information on Facebook do you think is available to the public?* (b) *What information do you think advertisers on Facebook already know about you?*

### 3.4 Discussion

In the experiment, we informed participants that the randomized price drawn from the computer was between $0 and $95. If a consumer's actual valuation is within this range, she has the incentive to report truthfully. If her valuation is above $95, reporting any value above $95 is optimal for her. In essence, it means that the reported values above $95 should be considered as "partially truthful": they truthfully reveal the fact that the underlying valuation is greater than $95, but are otherwise a stated preference.

Although many papers have used stated preference to measure privacy values (see Coyle & Manley 2022 for a review), debates abound over whether stated preferences are truthful (Spiekermann et al. 2001, Singleton & Harper 2002). One argument is that the gap between stated and revealed preferences diminishes once the context is controlled for (Prince & Wallsten 2020). Nevertheless, we adopt a variety of strategies in the data analysis to emphasize the valuations within the incentive-compatible range. The reduced-form analysis will focus on log valuation as our preferred specification. In addition, we will allow truncation at different finite points above $95 and see whether the results are sensitive to different specifications.

## 4 Data and Model Evidence

We recruited a total of 5,028 participants: 2,010 from Facebook during February 11-27, and 3,018 from Prolific during March 7-10, both in 2022. Table 1 provides summary statistics of our participants. Compared with the US representative demographics,[7] ours includes more females and are overall better educated; the distributions of age, income, and ethnic majority are similar to the national average. Compared to those recruited from Prolific, the Facebook ad participants include more females, are older, wealthier, better educated, more likely to be minorities, and spend more time on Facebook; they also click on ads and shop on Facebook more often. Thus, having participants from both sources allow us to cover a wider demographic range, which allows us to demonstrate how privacy valuations and frame effects differ across the demographic spectrum. Table B.2 and B.3 shows that our participants are balanced across the six treatment conditions.[8]

---
[7]https://www.census.gov/quickfacts/fact/table/US/PST045221; https://www.census.gov/library/visualizations/2022/comm/aging-nation-median-age.html.

Table 1: Summary Statistics of Participant Characteristics

|  | Overall | | Facebook | | Prolific | | U.S. Census |
| --- | --- | --- | --- | --- | --- | --- | --- |
|  | Mean | SD | Mean | SD | Mean | SD | Mean |
| **Number of participants** | | | | | | | |
| N | 5028 | | 2010 | | 3018 | | |
| **Race (percentage)** | | | | | | | |
| White | 0.8 | 0.4 | 0.73 | 0.44 | 0.85 | 0.35 | 0.76 |
| Black | 0.07 | 0.25 | 0.08 | 0.27 | 0.06 | 0.23 | 0.14 |
| Asian | 0.12 | 0.32 | 0.16 | 0.37 | 0.08 | 0.28 | 0.06 |
| Other | 0.05 | 0.05 | 0.06 | 0.06 | 0.04 | 0.04 | 0.05 |
| **Gender (percentage)** | | | | | | | |
| Female | 0.59 | 0.49 | 0.76 | 0.43 | 0.48 | 0.5 | 0.51 |
| **Median age** | | | | | | | |
| Median age | 39.5 | 12.77 | 39.5 | 13.23 | 39.5 | 12.22 | 38.8 |
| **Median household income ($)** | | | | | | | |
| Median household income | 62500 | 48675 | 62500 | 51629 | 62500 | 45224 | 64994 |
| **Education (percentage)** | | | | | | | |
| High school graduate or higher | 0.99 | 0.1 | 0.99 | 0.09 | 0.99 | 0.11 | 0.89 |
| Bachelor's degree or higher | 0.64 | 0.48 | 0.73 | 0.44 | 0.58 | 0.49 | 0.33 |
| **Facebook Questions** | | | | | | | |
| Average time spent on FB (h) | 1.41 | 1.38 | 1.97 | 1.37 | 1.03 | 1.25 | |
| FB membership duration (y) | 5.76 | 0.88 | 5.75 | 0.87 | 5.76 | 0.89 | |
| Average time spent on internet (h) | 3.52 | 1.84 | 3.5 | 1.82 | 3.54 | 1.86 | |
| Active user (percentage) | 0.31 | 0.46 | 0.47 | 0.5 | 0.21 | 0.41 | |
| Purchase from FB or Instagram (times/mo) | 0.41 | 0.78 | 0.63 | 0.95 | 0.27 | 0.6 | |
| FB or Instagram ad click (times/mo) | 1.66 | 1.81 | 2.47 | 1.92 | 1.12 | 1.51 | |

The second step in the WTA elicitation is unrestricted; thus participants can be inconsistent. For example, a participant may say they are willing to sell their posts for $40 but not for $30 in the MPL, then ask for $56 on the next screen. Appendix Figure B.1 shows responses from all participants who completed our study, with their free-text valuations on the Y-axis and the implied WTA from their MPL responses on the X-axis. We find that 93% of participants give consistent valuations throughout. Among participants who have given inconsistent answers, many deviate to a range close to where they were. More importantly, 83% of consumers who give inconsistent valuations only do so occasionally, suggesting that their inconsistency is more a sign of regret than a byproduct of inattentiveness.

---

[8]Here, we separate the covariate balance tests for the attributes measured in the baseline and endline surveys. One concern about the endline survey responses is that they may be influenced by the treatments, especially when it comes to beliefs about data usage. Table B.3 shows that the endline responses do not differ significantly across treatments and thus can be included in our heterogeneity analysis.

Consumers' valuations for their data are heavily skewed to the right: In fact, they report a value of infinity for 18.3% of the time. We allowed for infinity reporting because earlier work posits that some consumers can be "privacy fundamentalists" who would reject any benefits from data uses in exchange for privacy (Westin 2003, Woodruff et al. 2014), and we wanted the valuation measurement to allow for this possibility. Nevertheless, most consumers in our experiment do not seem to adopt a fundamentalist attitude in that they are selective when reporting infinite values. For example, 7% of participants report infinity valuation for their survey answers compared to 31.7% for friends and followers. Only 3.7% of participants report infinity values for all items.

Genuine or not, the infinity values create challenges in reporting summary statistics and reduced-form analysis. We adopt several strategies to account for this challenge. In the model-free evidence, we report most results in the form of data supply curves, with the percentage of consumers with infinite valuations at the top of each curve. Since the data supply curves are essentially cumulative distribution functions of the valuation, they transparently visualize the distribution of consumer valuations and the impact of choice frames on different parts of the curve. In the reduced-form models, we use log valuation as the outcome variable in our main specification and indicate where we top-code the data. Focusing on log valuations makes the results robust to the impacts of extremely high values and our top-code choices. We note that the long-tail pattern in privacy valuations is universal (e.g., see Collis et al. 2020 and Lin 2022) and is not unique to our setting.

## 4.1 Valuations Across Consumer Data

Our first result shows that consumers have systematically different valuations for different types of personal data (Figure 3). They value *friends and followers* data the most; next comes *posts* and *about me* information; *survey answers* are valued the least. When top-coded at $100, the mean valuation for *friends* data is $77.04; the average valuation across all Facebook data is $67.14, compared to an average of $48.64 for sharing the survey responses. Given that 20% of consumers report their data valuation at above $100, these value differences serve as the lower bounds for the actual differences.

The "coherent arbitrariness" theory (Ariely et al. 2003) argues that people can have coherent differences in valuation after choosing an arbitrary starting point. One may expect such a phenomenon to be likely for private data valuations, as the pros and cons of sharing data can be uncertain. To see if this hypothesis holds in our setting, we leverage the fact that the order of personal variables is also randomized across consumers and examine the valuations using only the first question each consumer encounters. The differences in valuation across data persist with a similar magnitude, even when we focus on only the first valuation (see Appendix B.2). As another test of choice coherence, we compare valuation for Posts and Likes data between participants who are randomized to different data duration conditions, and find that they assign a higher value to their online history data when it has a longer duration (see Appendix Figure B.3). These patterns show that consumers' valuation for privacy is not arbitrary, but coherent both within and across people.

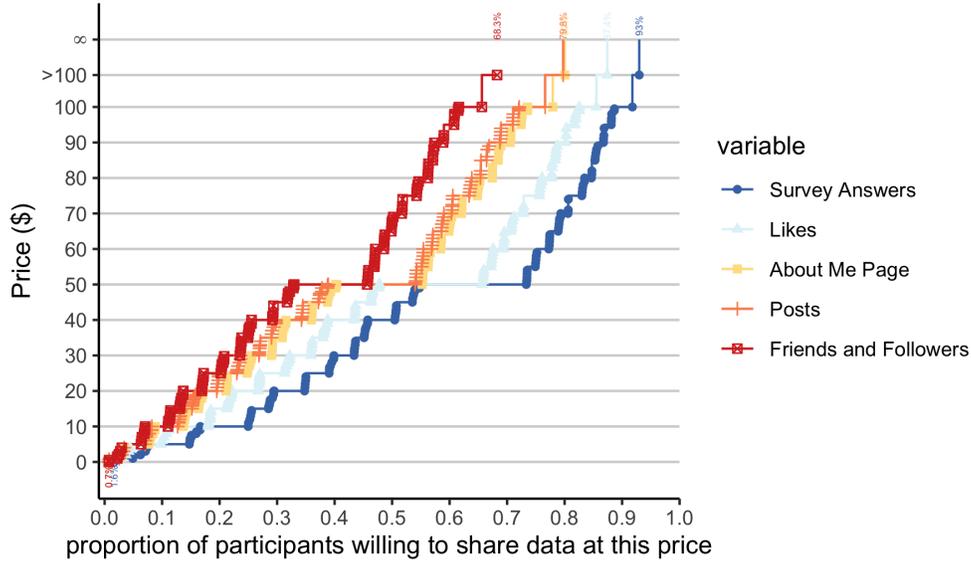

Figure 3: Data Supply Curves by Data Type

## 4.2 The Effects of Choice Architecture on Data Supply Curves

Despite the coherence, consumers' privacy valuations are prone to the influence of choice architecture. Table 2 summarizes the magnitude of average treatment effects using Tobit regressions, with the valuations top-coded at $100 as the outcome variable. Our preferred specification is Model 4, which uses the log form of valuations to decrease the sensitivity of estimates to top code choices, and includes the types of personal variables as additional controls. Compared to active choice, an opt-out frame decreases the average valuation by 5.8%, while opt-in increases the valuation by 7.8%. The influence of a price anchor is more substantial. Consumer valuations for data decrease by 52.6% on average when priced in the low-price as opposed to the high-price condition.[9]

Figure 4 further compares the data supply curve across all six treatment combinations. The default conditions shift the supply curve uniformly, with the supply curve corresponding to active choice sitting squarely between opt-in and opt-out. In contrast, the anchor price distorts different regions of the supply curve in different ways. In particular, the gap between supply curves is the largest in the middle region, due to valuations bunching towards the endpoints of the price range.

---

[9]For log outcome models, we transform the outcome using $\log(Y+1)$ instead of $\log(Y)$ to account for the presence of zero valuations. In our setting, $Y$ is often reasonably large, thus we directly read off the coefficients as percentage changes in the outcome induced by the treatments.

Table 2: Average Treatment Effects and Valuation Across Data: Tobit Regressions

|  | WTA | WTA | log(WTA) | log(WTA) |
|---|---|---|---|---|
| Intercept | 63.884 *** | 48.636 *** | 4.001 *** | 3.570 *** |
|  | (0.876) | (0.881) | (0.026) | (0.027) |
| Price Anchor = Low | -16.112 *** | -16.234 *** | -0.523 *** | -0.526 *** |
|  | (0.947) | (0.921) | (0.028) | (0.028) |
| Default = Active | 2.377 * | 2.205 * | 0.063 + | 0.058 + |
|  | (1.139) | (1.108) | (0.035) | (0.034) |
| Default = Opt-in | 5.178 *** | 5.101 *** | 0.138 *** | 0.136 *** |
|  | (1.147) | (1.119) | (0.035) | (0.034) |
| Likes |  | 8.269 *** |  | 0.268 *** |
|  |  | (0.455) |  | (0.014) |
| About Me Page |  | 17.875 *** |  | 0.509 *** |
|  |  | (0.523) |  | (0.015) |
| Posts |  | 19.459 *** |  | 0.550 *** |
|  |  | (0.535) |  | (0.016) |
| Friends and Followers |  | 28.399 *** |  | 0.764 *** |
|  |  | (0.628) |  | (0.018) |
| Num. Obs. | 25140 | 25140 | 25140 | 25140 |

+ p < 0.1, * p < 0.05, ** p < 0.01, *** p < 0.001. Outcomes variables are top-coded at $100; standard errors are clustered at the participant level. For the log outcome models, we transform the valuation using $\log(Y + 1)$ to account for the presence of zero valuations.

[10]Paradoxically, a low-price anchor also triggers more consumers to report extremely high (greater than $100) and infinite values. In the low-price condition, the average percentage of infinity values across variables is 21.2%, compared to 15.5% in the high-price condition. As a result, the supply curves from the two treatments intersect around the $95 price point, demonstrating the non-monotonicity of the price anchor effect.[11]

Table C.3 shows versions of the log outcome model with different top codes for the data valuation. As the top code becomes less stringent, the coefficient representing the price anchor effect decreases in magnitude while all other coefficients increase. One way to explain this finding is that price anchor creates "backlash" in a subset of consumers. Another potential explanation is that in the low price anchor treatment, consumers bother less with reporting a precise value because their data valuation is further away from the highest MPL price, and opt for infinity as a mental shortcut.

## 4.3 Heterogeneous Effects of Choice Architecture

In Section 2, we show that the correlation between consumers' privacy valuations in the neutral benchmark condition and their responses to choice frames is the key driver that creates the tension

---

[10]Although bunching is prevalent, it is not persistent within individual. In fact, only 3% of our participants choose their valuations at one of the price range endpoints persistently across five personal variables.

[11]Appendix Table C.2 includes further controls and variables of interest. Consumers who ask for more than the token's value in the practice round also report higher WTA for their data. In addition, consumers who believe specific data are already available to Facebook and to the public value their data less when it comes to sharing data with advertisers.

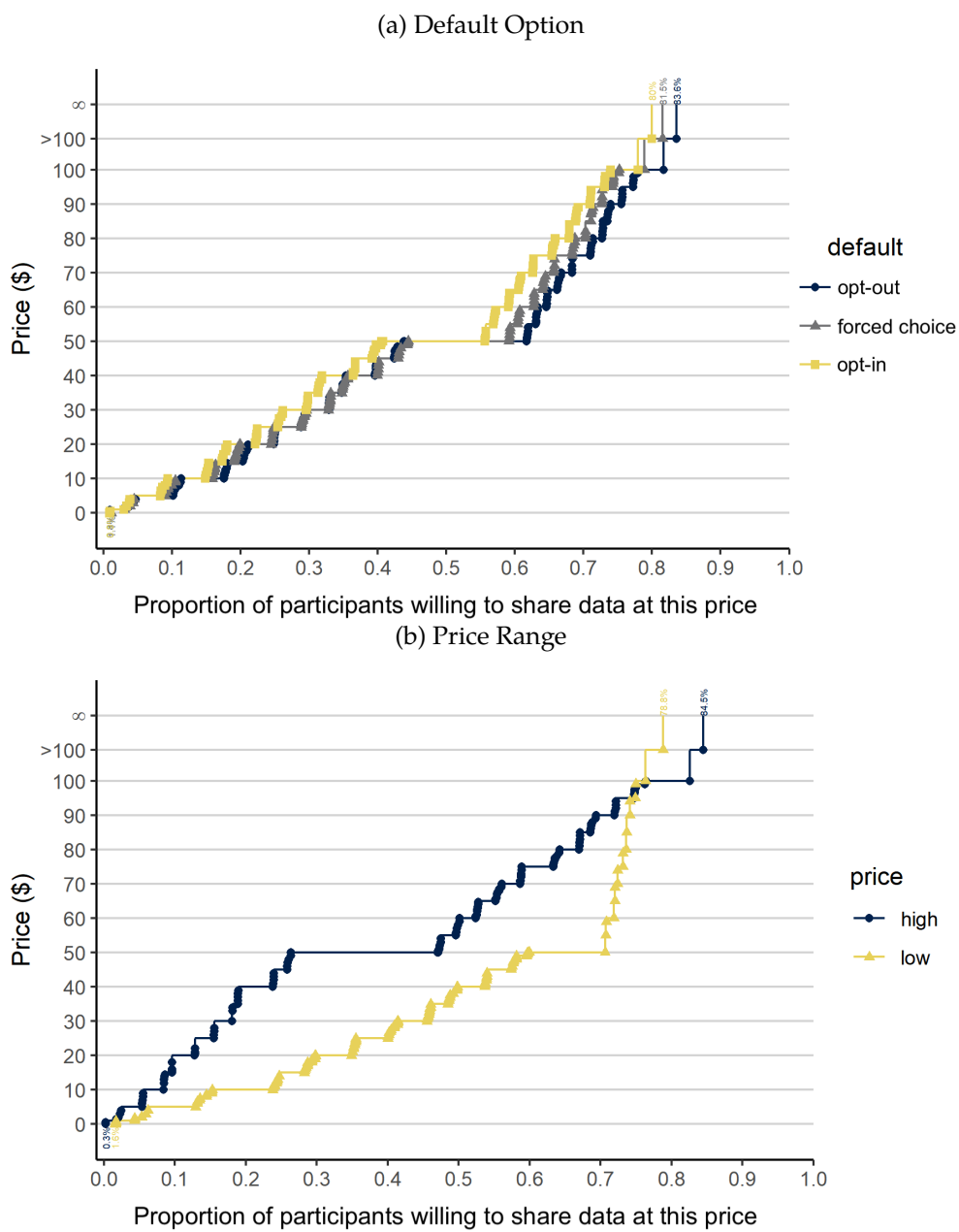

Figure 4: Data Supply Curves by Treatment Condition

(a) Default Option

(b) Price Range

between volume-maximizing and bias-minimizing effects across frames. Simply comparing data supply curves under different frames is insufficient for capturing such correlation, as the supply curves do not show how much the same consumer (segment) would value their data under different conditions. Instead, we must rely on heterogeneous effect models to characterize the joint distribution of consumers' privacy valuations and the frame effects.

To achieve this goal, we estimate causal forest models proposed by Athey et al. (2019). These models allow us to efficiently and flexibly capture the treatment effects across consumer sub-groups.[12] We adopt two specifications. Our preferred specification is a multi-arm forest, which captures potential interaction effects between default and price range treatments. The caveat of the multi-arm forest model is that it does not account for censored values (i.e., our top codes for infinite values). Our second specification is a survival forest, which takes care of censoring in a Tobit-model fashion, but only compares two treatments at a time and thus does not capture interaction effects. The heterogeneity patterns in the two models are qualitatively similar, though the results from the multi-arm forests are smaller in magnitude due to censoring. In what follows, we use results from the multi-arm forest for our analysis. We include details of model tuning and a comparison of estimation results across the two models in Appendix C.

For many choice frames including the price anchor, what counts as a "neutral" frame in practice is unclear. In view of this fact, we instead construct the benchmark privacy valuation as the average of a participant's would-be valuations across the six treatments, predicted by our model as $E[Y|X_i]$ (hereafter the *average* privacy valuation). This average valuation underpins a consumer's data-sharing choices when the firm randomly chooses among the six possible frames. It is possible that a true frame-neutral valuation lies closer to one of the frames or even outside this range, but this approach allows us to get a comparable valuation across participants without imposing strong parametric assumptions. In our main specification, we use log valuation as the outcome variable, where the valuations are top-coded at $1,000 before taking the log. The covariates in the model include demographics, Facebook and general internet usage, and consumer beliefs about what data are already available to Facebook as well as the public.

Figure 5 shows the distribution of the heterogeneous treatment effects estimates for the two frames. The average treatment effects for both frames are significantly above zero. Consistent with the raw data, the price range has a larger overall effect compared to the default. Together, the effects of opt-in and high price anchors increase WTA for almost all participants and vary in magnitude. Appendix C.1 shows the treatment effect estimates separately for each individual, along with standard errors.

Figure 6 shows the correlation between frame effects and the average log valuation. The effect of default does not vary systematically as the valuation increases (correlation is -0.04). On the other hand, the effect of price anchor is negatively correlated with the average valuation: consumers who have lower privacy valuations are also more likely swayed by a price anchor (correlation is -0.46). This pattern is consistent with Collis et al. (2020), who find that consumers who receive an informative signal on the data value often revise their valuation upwards when their prior is lower than the signal, but rarely update their belief downwards when their prior is high.

To further explore the overlap between consumers who have lower privacy valuations and those more responsive to choice frames, we project the average valuation and the heterogeneous "maximal frame effects" to the consumer attributes using linear models. The linear projection

---

[12]In comparison, simpler models such as linear regression do not have regularization built in and can perform poorly with a large number of consumer feature covariates.

Figure 5: Heterogeneous Treatment Effect Estimates: Multi-Arm Causal Forests

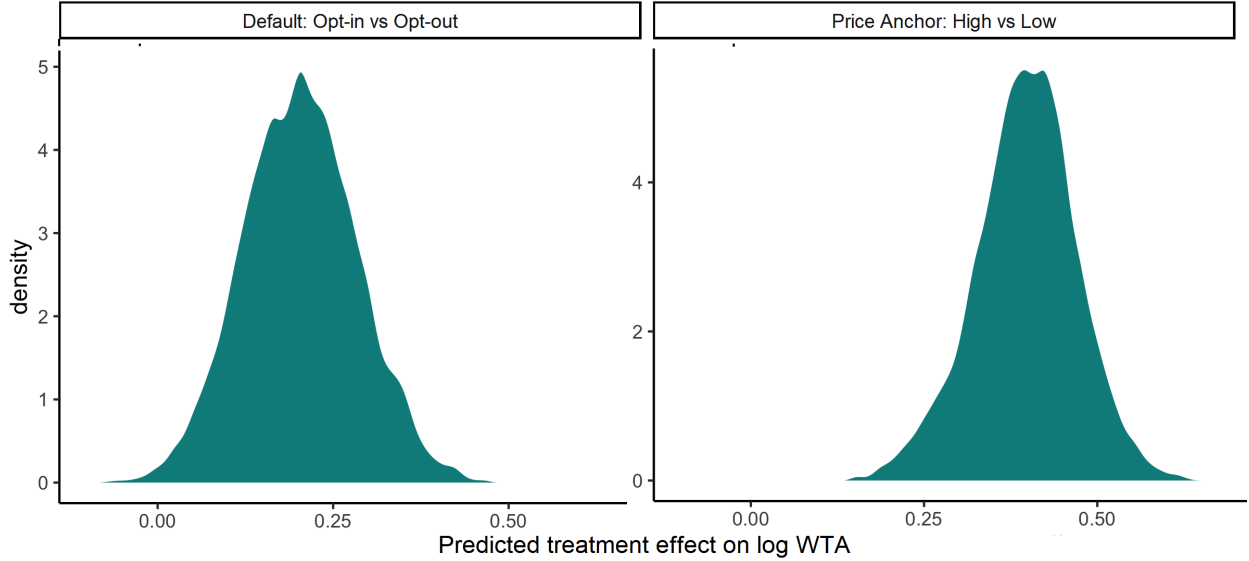

*Note:* Outcomes: log valuations truncated at $1000; standard errors are clustered at the subject ID level. ATE estimates from the multi-arm causal forest: $ATE_{\text{default}} = 0.09$ (se = 0.03); $ATE_{\text{price}} = 0.58$ (se = 0.03).

approach has been promoted by the recent literature (Semenova & Chernozhukov 2021) as an efficient way to summarize heterogeneity in a conditional expectation estimate. We define the maximal frame effect as the difference between the two frames that give the highest and lowest average treatment effects. In other words, we calculate for each consumer the difference in predicted valuation between the frame that maximizes the average reported valuation (opt-in, high price range) and the frame that minimizes it (opt-out, low price range).

Figure 7 shows the linear projection estimates and 95% confidence intervals. The top panel shows the predictors of a high baseline valuation. Consumers who value their personal data more are older, richer, better educated, more likely to be female and Asian, spend less time on the Internet, and are less likely to click ads on Facebook. In comparison, those more easily influenced by our choice frames are overall younger, poorer, less educated, and more likely to click on ads while using Facebook. Overall, these are attributes that predict high choice frame effects while also predicting lower valuations for personal data.

These heterogeneity results are broadly in line with findings in the existing literature. Regarding privacy valuations, Goldfarb & Tucker (2012) find that older people and women value their privacy more, and Collis et al. (2020) show that high-income consumers and Asian communities value their Facebook data more. As to frame effects, Mrkva et al. (2021) similarly find that lower socioeconomic status predicts stronger choice architecture effects. While we do not claim that our results generalize to all contexts, the consistency strengthens our belief that our results are robust.

Figure 6: Correlation Between Choice Frame Effects and Average Log Valuation

(a) Default Option

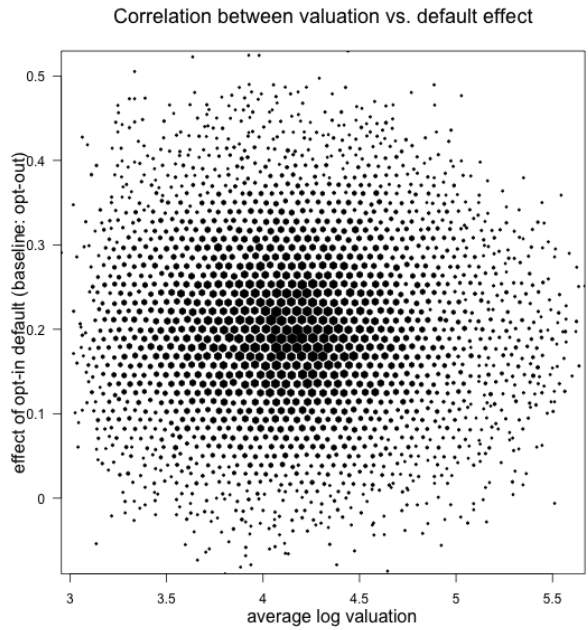

(b) Price Anchor

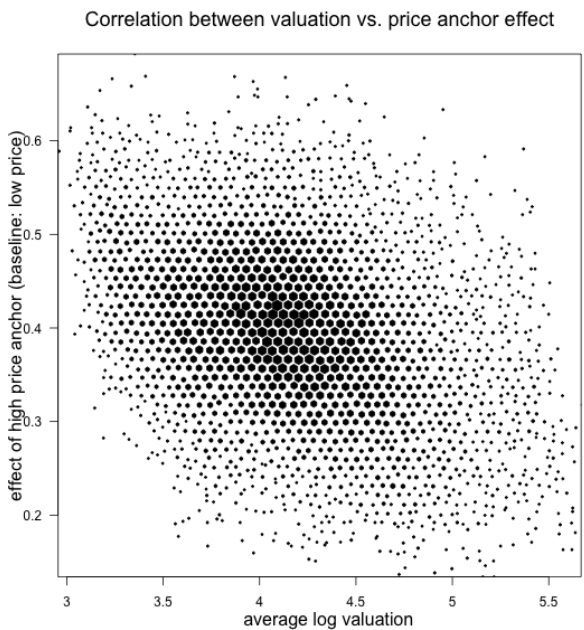

*Note:* In the figures above, treatment effects are estimated using multi-arm forests, with the outcome as log valuation truncated at $1000. The average valuations are generated alongside the model as the predicted outcome marginalized across the choice frames. The size of the points represents the number of observations in that region.

Figure 7: Heterogeneous Privacy Valuation and Treatment Effects by Consumer Subgroups

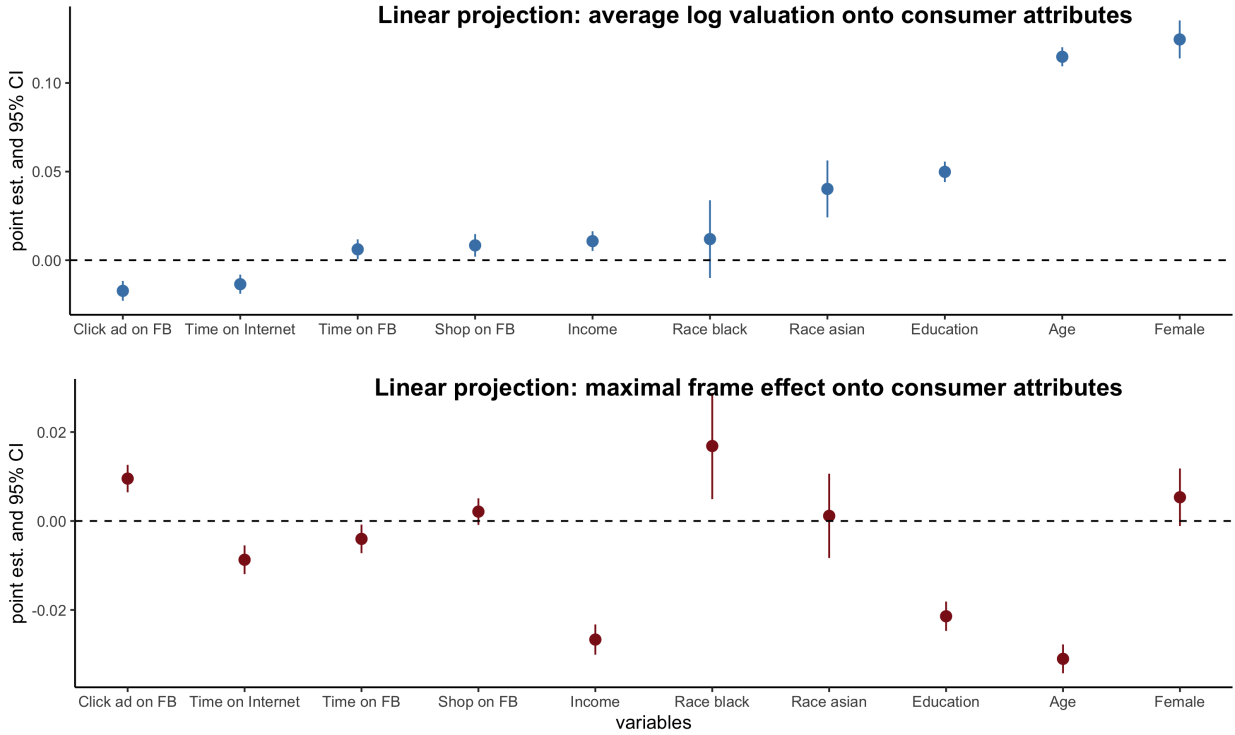

*Note:* In the first projection model, the outcome is a consumer's predicted log valuation averaged across the choice frame treatments. In the second projection model, the outcome is the predicted *maximal treatment effect*, represented as the difference between the choice frames with the highest and lowest ATE. Both outcome metrics come from the multi-arm causal forest estimates, with log valuation censored at $1000 as the outcome. All covariates are standardized before entering the model.

## 5 Exploring the Volume-Bias Trade-off

Our analysis has shown that choice frames can change the composition of consumers who share data since different consumers respond to frames differently—the question is how. With different frames, consumers have different willingness to share data. Does the frame chosen by the firm exacerbate or alleviate the existing selection bias? A company is often inclined to choose a choice architecture that maximizes the volume of data collected. Is this the optimal frame for data collection, or does volume come at the cost of representativeness?

To answer these questions, we conduct counterfactual simulations based on the raw data, leveraging the unconditional random assignment, and on our forest-model estimates. Our goal is to examine the volume and bias in data collected under different choice frames. In particular, we compare the performances of three frames: the *volume-maximizing* (hereafter *vol-max*) frame, which maximizes the supply of data at each price point; the *average* benchmark frame, which we construct by averaging a consumer's would-be valuations across all frames in our experiment; and the *bias-minimizing* (hereafter *bias-min*) frame, which minimizes the level of average bias at each price point. Our results show that a volume-maximizing choice frame can have opposite impacts on the bias in sample data through two distinct mechanisms. We also quantify the trade-off between the volume and representativeness objectives at the time of frame optimization, and show that the trade-off can change substantially depending on whether the firm is able to personalize its frame assignment.

## 5.1 When Does Bias in Data Degrade Its Value?

Before we proceed, it is worth clarifying what kind of data bias the firm cares about and when such bias decreases the value of consumer data. The value of data-driven analytics comes from learning and prediction. In other words, there is a series of outcomes $Y_i$ with distribution $f(Y|X)$ that the firm wants to learn about using consumer data. Examples of $Y$ include consumers' willingness to pay for certain products, interest in certain political topics or product categories, and price sensitivity. The firm observes $X$ (e.g., demographics) from all consumers at the point of data collection, but can only learn about $Y$ for consumers who agree to share their data. A biased sample means the sample distribution $f(X|\tilde{v}(\theta, X) > P)$ does not equal the target population distribution $f(X)$. Note that the target population can be the firm's desired customer database and need not represent the general population. Such a bias can compromise the statistical accuracy of data-driven insights and degrade the value of shared consumer data in the following scenarios:

1. The firm cares about learning the average outcome $E[Y] = E[f(Y|X) \cdot f(X)]$. However, the sample data gives $f(Y|X) \cdot f(X|\tilde{v}(\theta, X) > P) = f(Y|\tilde{v}(\theta, X) > P)$. When the firm does not know either $f(X)$ or $f(X|\tilde{v}(\theta, X) > P)$, it is unable to recover $E[Y]$ by reweighting the sample data to match the target population distribution.[13] This is represented by the *Product Hunt* example (Cao et al. 2021), where startups knew their target customer base but had no idea about the gender (or other demographic) representation of the votes from the platform.

2. The firm cares about the full, heterogeneous distribution of the outcome $f(Y|X)$ and is able to observe $X$ from the sample data. That allows them to learn $f(Y|X)$; however, some consumers $X = x_1$ are underrepresented in the dataset, which means that the accuracy of estimate for $f(Y|X = x_1)$ is low. This is typical in settings where the firm (or other data users) needs to know the heterogeneity of outcomes to design targeted campaigns and interventions.

In theory, $f(Y|X)$ may not always differ across $X$'s (i.e., no systematic relationships between the outcome and observable characteristics). However, firms often need to use the same consumer data to learn about many different outcomes, making this possibility exceedingly low in practice. In a typical data market, the data buying firm is an intermediary (e.g., Meta, Google, Experian) that later resells the raw or derived data to different end-user firms.[14] Since different end-user firms have different customer bases and different analytic objectives, the data-buying firm has to consider all these use cases. Even in cases where the data-buying and end-user firms are one and the same, the firm still needs to form an expectation over potential future uses of data, hence the phrase "data is a strategic asset". In situations like these, the objective of minimizing biases in the learned outcome vector $Y$ boils down to the objective of minimizing biases in observed attributes $X$ in the sample data, while the firm remains agnostic about the specific data production in each use case. Motivated by this observation, we focus on the representativeness of observables $X$ in the sample data in our following analysis.

---
[13]Reweighting is not a panacea either, as it typically decreases the effective sample size (ESS). The ESS decreases more as the variance of the weights increases (Stantcheva 2022), which happens when the sample data is more biased compared to the target population distribution. This fact again shows that more biases in the sample data can compromise the benefits of having a larger dataset.

## 5.2 Setup

We use the raw data to construct the data supply curves under each choice frame: Consumers with data valuations lower than the offer price are those who share their data.[15] These supply curves allow us to figure out who will share data for a given price and frame, thereby constructing the *sample data* under different price points and frames. We repeat the sample data calculation for a grid of prices between $20 and $90, with $1 increments.[16] For each sample data, we calculate its bias and volume:

- *Bias* is represented as the standardized difference in attributes between consumers who share data and all consumers in our experiment. For continuous attributes such as income, this is the standardized difference in means; for discrete attributes, this is the difference in the percentage of consumers with the attribute. We first focus on biases in individual attributes to illustrate the different mechanisms, then average over the standardized mean difference to form an aggregate bias measure.

- *Volume* is represented as the percentage of consumers captured in the sample data. Although alternative metrics for volume exist, using the percentage metric allows us to better characterizes how the volume-maximizing frame changes the bias in data in different regions.

We then identify three types of choice frames and compare the quality of data collected under each frame. In our setting, the uniform volume-maximizing frame means opt-out default and low price anchor, while the bias-minimizing frame is found by searching across all six frames for each given price. To construct valuations and choices under the average frame, we average the valuations across the frames in our experiment.

---

[14]Selling derived data can mean selling impressions from certain consumer segments, where the advertisers specify what segments they want to reach based on demographics or inferred consumer interests.

[15]When constructing the supply curve, we averaged each consumer's valuations across 5 personal variables, as we do not focus on the distinction across personal variables in this counterfactual.

[16]We choose the range of price grid such that the lowest price still guarantees that a nonzero percentage of consumers will share data in the average frame. If the percentage of consumers sharing the data is zero, we will not be able to calculate the bias metric. The highest price is more flexible, though due to the long-tail nature of consumer valuation, increasing prices in $1 increments at the higher end does not move the amount of data shared as much, thus we stop at $90.

## 5.3 The Bias and Volume of Data Across Frames

### 5.3.1 How Volume-Maximizing Frame Affects Data Bias: Two Countervailing Mechanisms

We first compare the data quality under the vol-max frame and the average benchmark frame under two extreme cases of market equilibrium. The first case compares the data quality when the firm's demand for data is perfectly inelastic, while the second represents the case when the firm has a perfectly elastic demand. These two cases serve as "bookends" that bound the general market outcomes, where the firm has a downward-sloping demand curve. Below, We illustrate the bias-variance trade-off in dimensions where a negative correlation between the privacy valuation and frame response is present: These are *income*, *age*, and *education*, as identified by our forest model estimates. We then expand the trade-off comparison by including all demographic covariates in the bias metric.

Figure 8 shows the biases and volumes of the sample data collected under the vol-max and average frames. We first focus on the case when the firm has perfectly inelastic demand. For instance, the firm may have a sample size target to run an experiment or construct a target segment, after which additional data has almost zero marginal value. To this end, we compare the bias for each pair of sample data with the same vertical position (thus having equal sample sizes) collected under each frame. We see that data collected under the volume-maximizing frame are generally further away from the representative benchmark (the solid line at zero), meaning they are more biased. This pattern confirms our intuition: When consumer segments initially with low valuations for data are influenced by the frame more, they become much more willing to share data under the frame, and thus the sample data over-represents these consumers.

However, Figure 8 also reveals a positive relationship between volume and representativeness within each frame: as the percentage of consumers sharing data increases, the sample becomes less biased overall. As an example, imagine the first 50% of consumers who share the data are all low-income consumers, while the next 50% are high-income ones. As the coverage goes beyond 50%, the sample eventually becomes less biased, because all the low-income consumers already choose to share data, and the new consumers at the margin serve to mitigate the sample bias. In other words, although the vol-max frame can exacerbate the bias when there is a negative correlation between privacy values and frame effects (Mechanism 1), it can also mitigate the bias, since it gets a higher coverage of consumers for a given price (Mechanism 2).

This observation implies that the trade-off can shift towards favoring the vol-max frame when the firm's demand for data is elastic. To compare data quality when the firm has perfectly elastic demand, we compare every pair of data collected under the same price but different frames, connected by the gray lines. Under the fixed-price comparison, a gray line with direction pointing from top-left to bottom-right implies that the vol-max frame alleviates bias, while a line pointing from top-right to bottom left indicates the opposite. The two mechanisms now counteract each

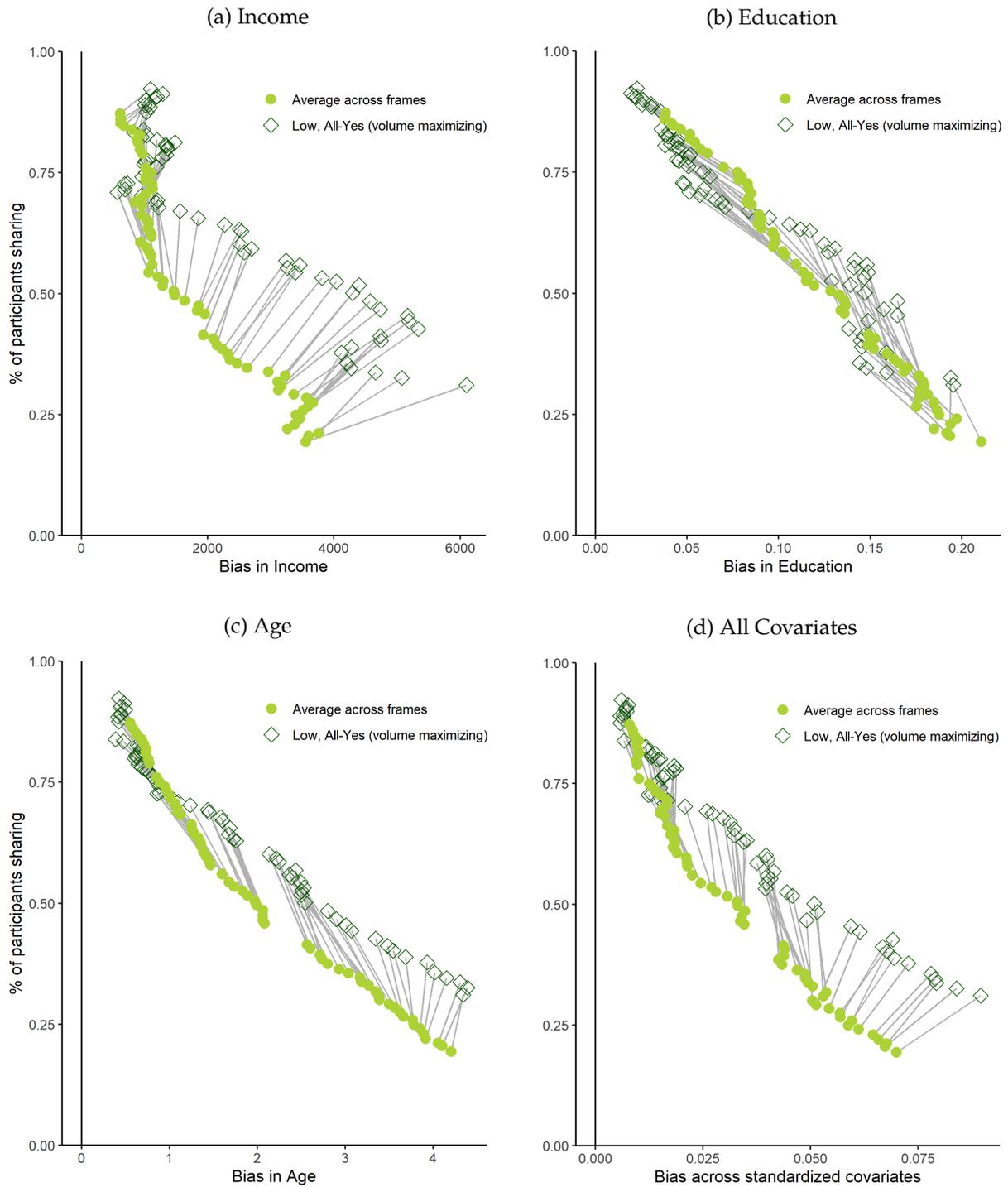

Figure 8: Data Quality Under Vol-Max and Average Frames

*Note:* In the figures above, each point represents a sample dataset collected under a choice frame × price combination. The x-axis represents the bias—the difference in mean attribute values between the sample and the full data, and the y-axis represents the size of the dataset. The prices for data vary from $25 to $90 in $1 increments; points connected by a gray line are collected under the same price but different frames.

other: While the bias-exacerbating effect can dominate when the coverage of data is low, the bias-mitigating effect starts to take over as the coverage of data increases.

### 5.3.2 Quantifying the Volume-Bias Trade-off

How may the firm choose the frames differently if it also cares about data representativeness? Different firms may prioritize representativeness or volume differently, depending on how they use data to improve their business. To characterize the volume-bias trade-off, here again we compare the optimal frame choices under two extreme preferences: volume maximization and bias minimization. Figure 9 compares the performances of these two frames conditional on the same price for data. The gap between vol-max and bias-min frames is large when the sample size is small, whereas the two frames converge as the sample size approaches 100%. This pattern suggests that the trade-off between the two objectives is more drastic when the firm is still at the initial stage of gathering consumer data, which is more often the case for new entrants and smaller firms. In comparison, larger and more established firms are more likely to find that volume-maximizing frames also effectively reduce bias when their demand for data is elastic.

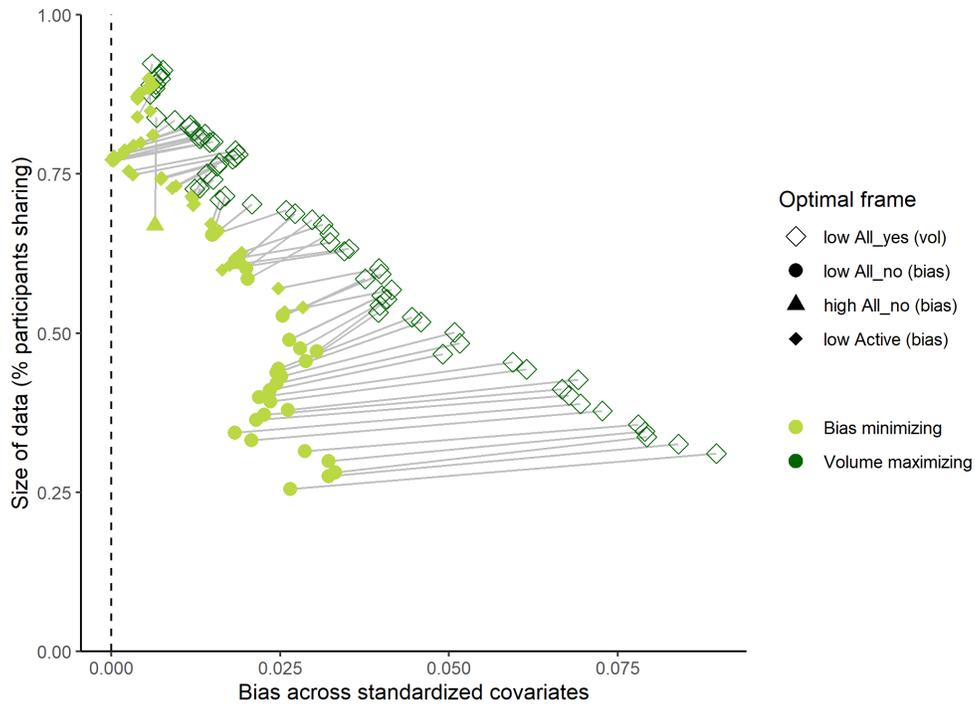

Figure 9: Data Quality Under Bias-Min and Vol-Max Frames: All Demographics Data

*Note:* In the figures above, each point represents a sample dataset collected under a choice frame × price combination. The x-axis represents the bias—the difference in mean attribute values between the sample and the full data, and the y-axis represents the size of the dataset. The prices for data vary from $25 to $90 in $1 increments; points connected by a gray line are collected under the same prices but different frames.

Figure 9 also shows that in our setting, the firm can sacrifice relatively little volume to achieve substantial gains in bias reduction. This is evident from the fact that moving from the vol-max to the bias-min frame while holding the price fixed moves the sample data substantially closer to the representative benchmark while moving down only slightly.

We create several metrics to quantify the volume-bias trade-offs as the firm optimizes its frame. First, we calculate the relative bias reduction and the relative volume reduction as the firm moves from the vol-max to the bias-min frame. Here, we define bias by comparing the average of standardized consumer covariates between the sample and the full data. The bias reduction takes the value of 0 if the two frames coincide, and 1 if all bias is eliminated; the volume reduction metric is defined similarly. We then define two elasticities: The volume-to-bias elasticity is the ratio between the percentage decrease in volume and the percentage decrease in bias as the firm moves from the vol-max to the bias-min frame, holding the price for data fixed. We also define the volume-to-price elasticity, which is the ratio between the percentage increase in volume and the percentage increase in price, holding the frame fixed.

The trade-off can depend on whether the firm can customize the frame for each consumer or is constrained to uniform frame assignment. For example, sometimes the firm may already know the demographic attributes of the consumers before asking them to share their behavioral data. In view of this possibility, we calculate these trade-off metrics separately for the uniform and personalized frame assignment and across different price points. To demonstrate the impact of personalization while sticking to the purely model-free visualization scheme, we divide the participants into above and below median covariate index level, and search through the 36 potential assignments. We provide the summary statistics of the trade-off metrics in Figure 10 and the full set of metric values in Appendix C.4.

With uniform frame assignment, moving from the vol-max frame to the bias-min frame leads to a 46% reduction in average bias and an 8% loss of data volume. The volume-to-bias elasticity varies a lot across prices, with a median of 0.17 and an average of 0.85. These numbers imply that a 1.7%-8.5% decrease in volume accompanies a 10% decrease in bias. The elasticity is often smaller than 1, consistent with our previous finding that a smaller volume reduction is needed for a larger bias reduction gain. The volume-to-price elasticity has a median of 0.7 and a mean of 0.79. To understand how much the price needs to increase to offset the volume loss caused by bias-reducing frames, we can calculate the rate of substitution by dividing the two elasticities.[17] The rate of substitution is 0.24 when calculated using median measures and 1.08 under the average, meaning the firm needs to increase the price for data by 2.4-10.8% for a 10% reduction in bias if it does not want to sacrifice volume.

The magnitudes of these trade-off metrics change substantially when the firm is able to customize frames across consumers. With personalized frame assignment, moving from vol-max

---

[17]This is a crude measure of substitution, as the volume-to-price elasticity holds the frame and not the bias as fixed.

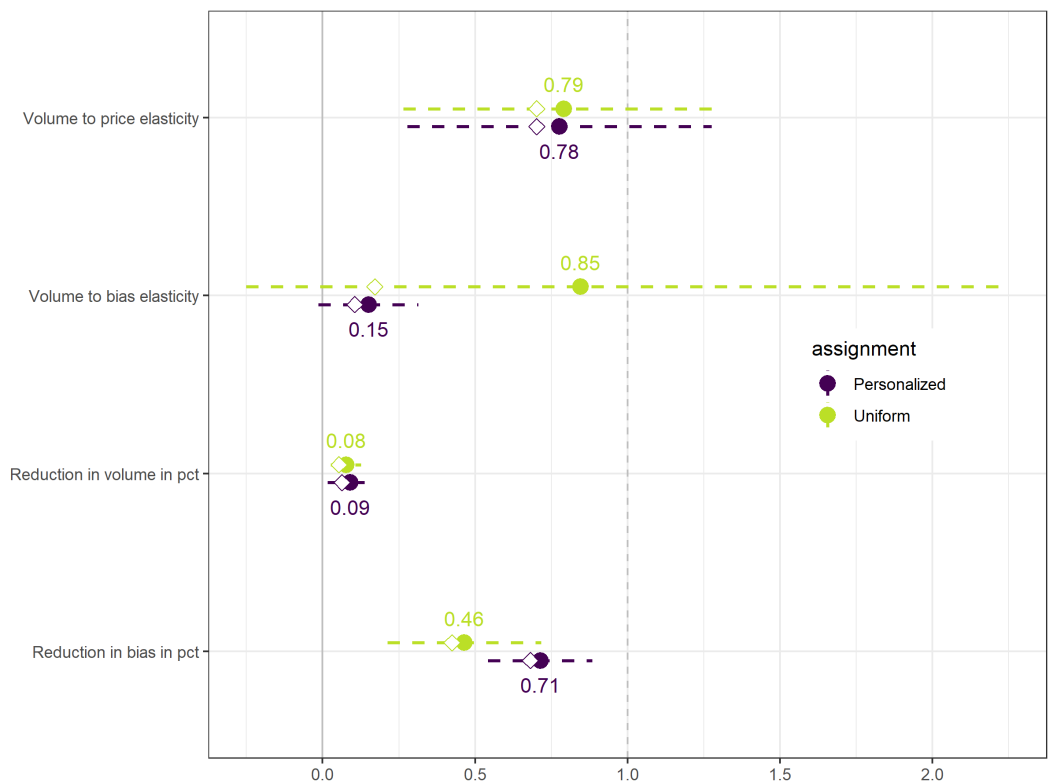

Figure 10: Summary Statistics of Bias-Variance Trade-off Metrics

*Note:* The figure shows summary statistics of trade-off metrics averaged across price points. The rounded points correspond to the mean, and diamond-shaped points correspond to medians. The length of the horizontal lines represents one standard deviation. The green and purple colors represent the values corresponding to the uniform and personalized frame assignment, respectively.

to bias-min leads to a more substantial bias reduction gain of 71%, almost twice as much as the gain in the uniform frame regime. In comparison, the reduction in volume is small and similar between the two regimes (9% versus 8%). These numbers translate to a lower volume-to-bias elasticity of 0.1-0.15. The small loss in volume may come from the fact that the most effective frame in shifting down consumers' valuations is mostly the same across consumers (opt-out and low-anchor) in our setting. On the other hand, balancing consumer composition requires more granular control and can generally be improved with frame personalization.

How can the firm choose the optimal frame in practice? The key to optimizing the frame for data collection lies in knowing the joint heterogeneity of privacy preferences and choice frame responses. We propose a sequential optimization strategy, where the firm uses contextual bandit to simultaneously learn about this joint heterogeneity and choose the optimal frame given its current knowledge. Using a multi-armed bandit to search for the optimal design is an existing feature in several experiment-as-a-service platforms.[18] Alternatively, firms may opt for a learn-then-earn framework to achieve the same goal.

---

[18]See: https://vwo.com/blog/multi-armed-bandit-algorithm/, and https://www.optimizely.com/optimization-glossary/multi-armed-bandit/.

# 6   Conclusion

Choice architecture can substantially distort consumers' privacy valuations and thus the supply of consumer data. The choice frames we examine—default and price anchor—shift the average consumer valuation for their Facebook data by 14% and 53%, respectively. Moreover, for some consumer segments, the susceptibility to frame influence is negatively correlated with their valuation for data absent frames. Younger, lower-income, and less educated consumers tend to respond more strongly to changes in the frame; they also value their personal data less on average.

The fact that different consumers respond to frames differently implies that choice frames can substantially change the composition of data that firms collect. We show that a conventional practice—choosing a frame that maximizes the supply of data—can have opposite effects on the bias in the data collected. In cases where consumers who value data less also respond more to choice architecture, a frame that aims to decrease valuations and maximize data supply can exacerbate the bias in data. However, a frame that maximizes the supply of data can also mitigate the bias, as increasing the volume also means eventually improving the coverage of the sample data. We show that the bias mitigating effect tends to dominate when the firm can get a high percentage of consumers to share their data by providing a high price. We also show how alternative frame designs can allow us to achieve substantial gains in bias reduction without sacrificing much volume, especially when frames can be personalized.

The correlational pattern between privacy preferences and frame effects may vary across economic contexts due to consumers' instrumental preference for privacy changing (Lin 2022). We do not claim that the heterogeneity patterns in our setting generalize to other contexts where firms collect consumer data for purposes different from advertising. Rather, we aim to provide an empirical example to show the mechanisms under which choice architecture can create tension between the two data quality objectives.

Our results contribute to the broad discussions about the efficiency of the data market. Several features unique to consumer data can lead to market failures, such as externalities (Bergemann et al. 2022), incomplete information (Jin 2018), and non-rivalry (Jones & Tonetti 2020). Here, we have shown another feature that contributes to market inefficiency: a data market that allows consumers to make their own choices can often create bias in data collected and decrease the returns to data. In the context of choice architecture evaluation, we show when companies' efforts to maximize data collection can exacerbate or alleviate the bias in different market conditions. However, a data collector attuned to reducing bias can leverage smart choice architecture as a tool rather than a hindrance. We believe the bias aspect is worth emphasizing, as it represents a novel channel in the data market that affects its functioning.

# A   Experiment Appendix

Figure A.1: Example Recruiting Ad

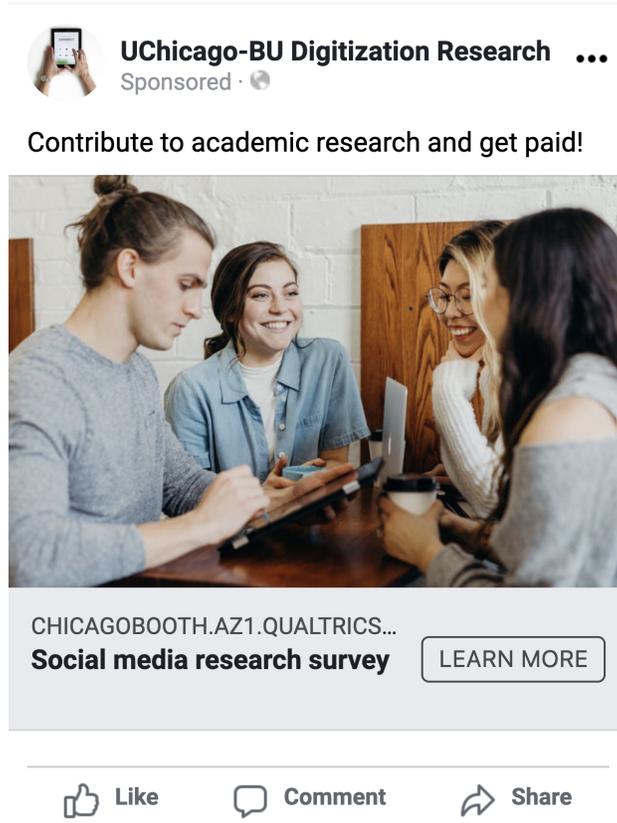

Figure A.2: Information Prompts in the MPL Practice Question

(a) Accepting an offer price too low

⚠ If you choose less than $14.50, you might be selling the pen less than what's worth for you.

**Will you sell your pen for $14?**

🔘 Yes

⚪ No

(b) Rejecting an offer price too high

⚠ If you choose not to sell for a price more than $14.50, you might not sell the pen even though you want to.

**Will you sell your pen for $16?**

⚪ Yes

🔘 No

Figure A.3: The Multiple Choice List Procedure: Step 1

In this question, we are asking for your price to share **Posts:** your Facebook posts and feed history from the last month.

**If the computer chose any of the following prices, will you share your data?**

|  | Your choice | |
|---|:---:|:---:|
|  | Yes | No |
| Will you share your **Posts** for $50? | ○ | ○ |
| Will you share your **Posts** for $60? | ○ | ○ |
| Will you share your **Posts** for $70? | ○ | ○ |
| Will you share your **Posts** for $80? | ○ | ○ |
| Will you share your **Posts** for $90? | ○ | ○ |
| Will you share your **Posts** for $100? | ○ | ○ |

*Note:* This screenshot shows what a participant sees in the first step of our MPL procedure in the *high price range + forced choice* condition. In the opt-in condition, all the "yes" options are pre-selected; in the opt-out condition, the "no" options are pre-selected. If a participant gets randomized into the *low price range* condition, the prices listed will range from $0 to $50, instead of $50 to $100.

Figure A.4: The Multiple Choice List Procedure: Step 2

(a) Scenario 1: participant says "yes" to some prices but not all

You said you are willing to share your Posts for $90 but not for $80, is there a more exact price that you are willing to share the data for?

→

(b) Scenario 2: participant says "no" to all prices listed

You have indicated that you will not share Posts for prices listed on the previous page. Can you tell us why?

○ The price I am willing to share Posts was not listed. I am willing to share it for (please type the desired price without $)

○ I do not want to share Posts for any price

*Note:* The screenshots above show two examples of what a participant sees in the second step of our MPL procedure. In the first scenario, a participant agrees to share their posts data for $90 or above but not for $80 or below. In the second scenario, a participant chooses not to share their *posts* data for all the prices listed in the first step. In this case, we give them the option to indicate that their valuation for the *posts* data is infinity.

# B  Data Appendix

Table B.1: Attrition throughout the Study

|  | Facebook | Prolific |
| --- | --- | --- |
| Exposed to the ad | 158453 | - |
| Click on the ad | 10135 | - |
| Consent to participate | 2348 | 3119 |
| Complete all questions | 2008 | 3018 |

Table B.2: Covariate Balance Across Treatments: Baseline Variables

| | Treatment | | | | | | T-test stat | |
|---|---|---|---|---|---|---|---|---|
| | low: opt-out | low: active | low: opt-in | high: opt-out | high: active | high: opt-in | | p-value |
| **Number of participants** | | | | | | | | |
| n | 786 | 824 | 817 | 832 | 815 | 831 | | 0.321 |
| **Race (percentage)** | | | | | | | | |
| White | 0.838 | 0.8 | 0.819 | 0.802 | 0.801 | 0.812 | 0.457 | |
| Black | 0.055 | 0.073 | 0.059 | 0.073 | 0.071 | 0.059 | 0.773 | |
| Asian | 0.106 | 0.112 | 0.105 | 0.108 | 0.118 | 0.125 | 0.42 | |
| Other | 0.034 | 0.045 | 0.049 | 0.056 | 0.052 | 0.047 | 0.012 | |
| **Gender (percentage)** | | | | | | | | |
| Female | 0.592 | 0.586 | 0.595 | 0.594 | 0.574 | 0.592 | 0.191 | |
| **Age** | | | | | | | | |
| Age | 39.398 | 37.885 | 38.359 | 38.864 | 39.124 | 38.526 | 0.146 | |
| **Income ($)** | | | | | | | | |
| Income | 66724 | 67433 | 72080 | 70354 | 73156 | 66347 | 0.962 | |
| **Education (percentage)** | | | | | | | | |
| College education and above | 0.644 | 0.623 | 0.647 | 0.637 | 0.659 | 0.643 | 0.87 | |
| Some college | 0.19 | 0.184 | 0.177 | 0.198 | 0.179 | 0.182 | | |
| Less than college | 0.167 | 0.193 | 0.175 | 0.165 | 0.162 | 0.176 | | |
| **Facebook and Internet usage** | | | | | | | | |
| Avg time spent on FB (h) | -0.051 | 0.061 | 0.01 | -0.046 | 0.039 | -0.027 | 0.755 | |
| FB membership duration (y) | 0.037 | 0.028 | 0.095 | 0.019 | 0.023 | 0.014 | 0.712 | |
| Avg time spent on internet (h) | 0.009 | 0.054 | -0.013 | -0.021 | 0.002 | -0.016 | 0.872 | |
| Active user (percentage) | 0.298 | 0.301 | 0.296 | 0.321 | 0.326 | 0.289 | 0.236 | |
| Purchase from FB or Instagram (times/mo) | -0.023 | -0.032 | 0.003 | -0.012 | -0.02 | -0.065 | 0.365 | |
| FB or Instagram ad click (times/mo) | -0.007 | -0.024 | 0.01 | -0.024 | -0.002 | 0.042 | 0.019 | |

Table B.3: Covariate Balance Across Treatments: Endline Variables

| | Treatment | | | | | | T-test stat |
|---|---|---|---|---|---|---|---|
| | low: opt-out | low: active | low: opt-in | high: opt-out | high: active | high: opt-in | p-value |
| **Number of participants** | | | | | | | |
| n | 786 | 824 | 817 | 832 | 815 | 831 | 0.321 |
| **Information available to the public** | | | | | | | |
| Info from the about page | 0.692 | 0.704 | 0.689 | 0.702 | 0.714 | 0.745 | 0.457 |
| Posts | 0.375 | 0.396 | 0.381 | 0.375 | 0.404 | 0.397 | 0.773 |
| Photos | 0.427 | 0.428 | 0.414 | 0.401 | 0.415 | 0.438 | 0.42 |
| Lists of likes | 0.309 | 0.317 | 0.321 | 0.312 | 0.324 | 0.337 | 0.012 |
| Friends and followers | 0.469 | 0.527 | 0.504 | 0.493 | 0.482 | 0.51 | 0.962 |
| Don't know | 0.162 | 0.126 | 0.151 | 0.16 | 0.144 | 0.144 | 0.191 |
| Other information | 0.085 | 0.104 | 0.075 | 0.079 | 0.106 | 0.066 | 0.146 |
| **Information available to Facebook advertisers** | | | | | | | |
| Name and email | 0.725 | 0.718 | 0.704 | 0.69 | 0.733 | 0.729 | 0.755 |
| Info from the about page | 0.822 | 0.816 | 0.815 | 0.831 | 0.847 | 0.836 | 0.712 |
| Posts | 0.531 | 0.535 | 0.526 | 0.507 | 0.531 | 0.521 | 0.872 |
| Photos | 0.468 | 0.45 | 0.454 | 0.425 | 0.456 | 0.452 | 0.236 |
| Lists of likes | 0.711 | 0.708 | 0.71 | 0.69 | 0.739 | 0.71 | 0.365 |
| Friends and followers | 0.585 | 0.631 | 0.624 | 0.588 | 0.632 | 0.611 | 0.019 |
| Don't know | 0.109 | 0.103 | 0.115 | 0.105 | 0.109 | 0.105 | 0.353 |
| Other information | 0.038 | 0.038 | 0.037 | 0.035 | 0.022 | 0.022 | 0.484 |
| **Participants looking up information** | | | | | | | |
| Looked up information when aswering questions | 0.019 | 0.023 | 0.021 | 0.023 | 0.028 | 0.031 | 0.891 |
| **The type of information participants look up** | | | | | | | |
| How advertisers use data for targetting | 0.008 | 0.007 | 0.004 | 0.01 | 0.011 | 0.011 | 0.659 |
| What Facebook shares with advertisers | 0.006 | 0.01 | 0.011 | 0.013 | 0.017 | 0.011 | 0.429 |
| How much each data is worth | 0.003 | 0.004 | 0.006 | 0.006 | 0.01 | 0.01 | 0.182 |
| Other information | 0.01 | 0.012 | 0.006 | 0.008 | 0.007 | 0.017 | 0.974 |
| Frequency of encountering cookie banners/day | 1.523 | 1.657 | 1.575 | 1.561 | 1.609 | 1.552 | 0.15 |

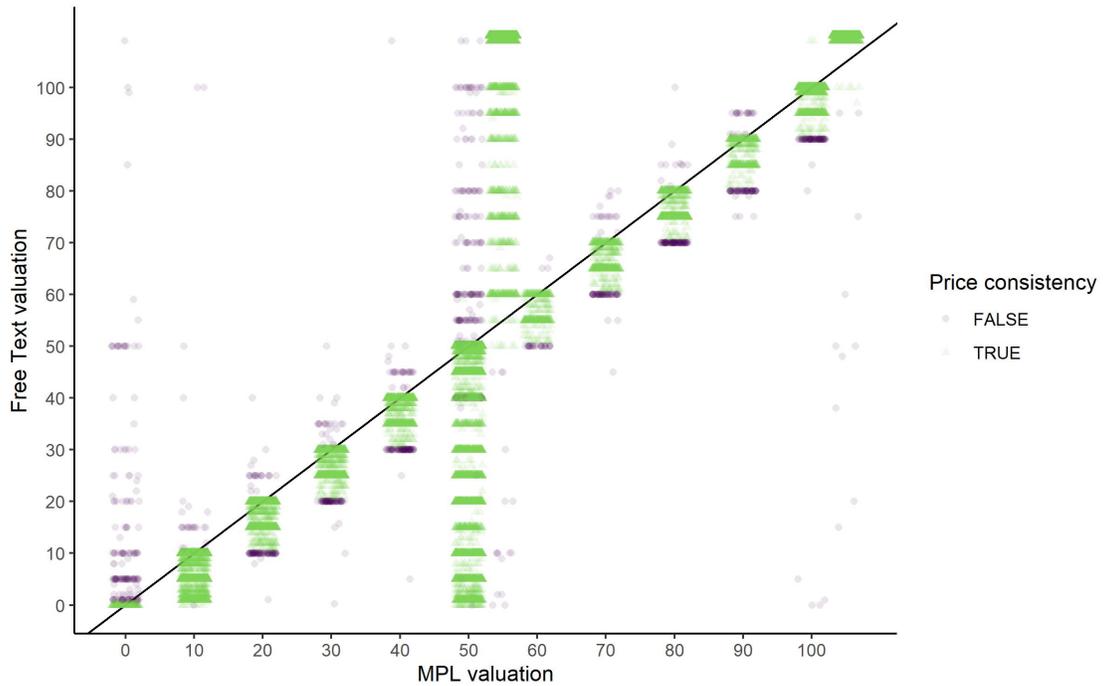

Figure B.1: Consistency Between Final Price and MPL Choice

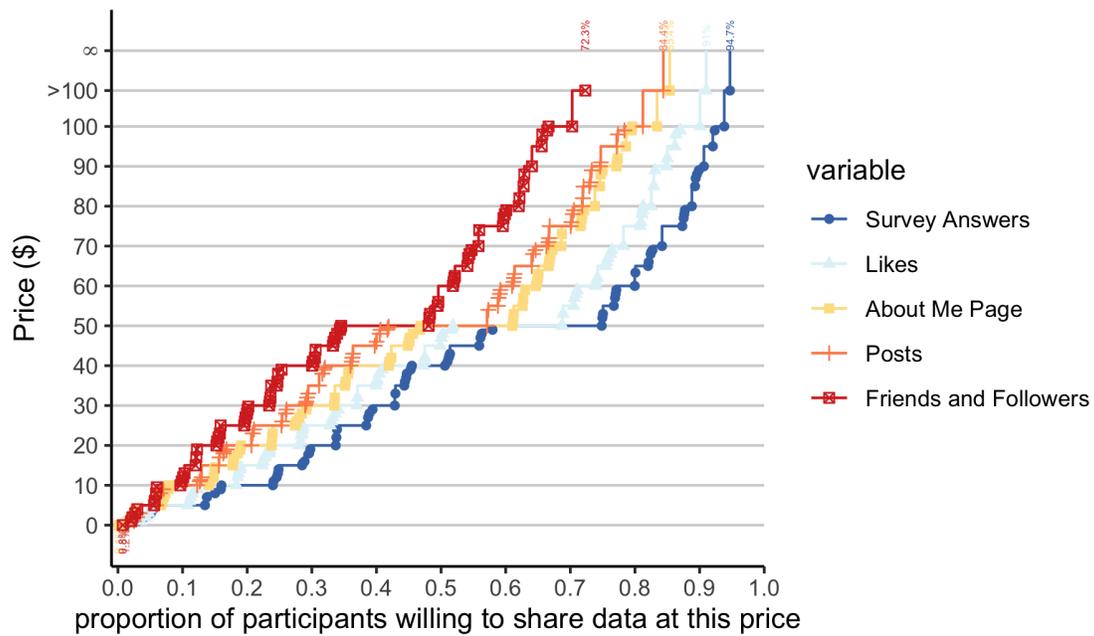

Figure B.2: Data Supply Curves by Data Type (First-Round Answer Only)

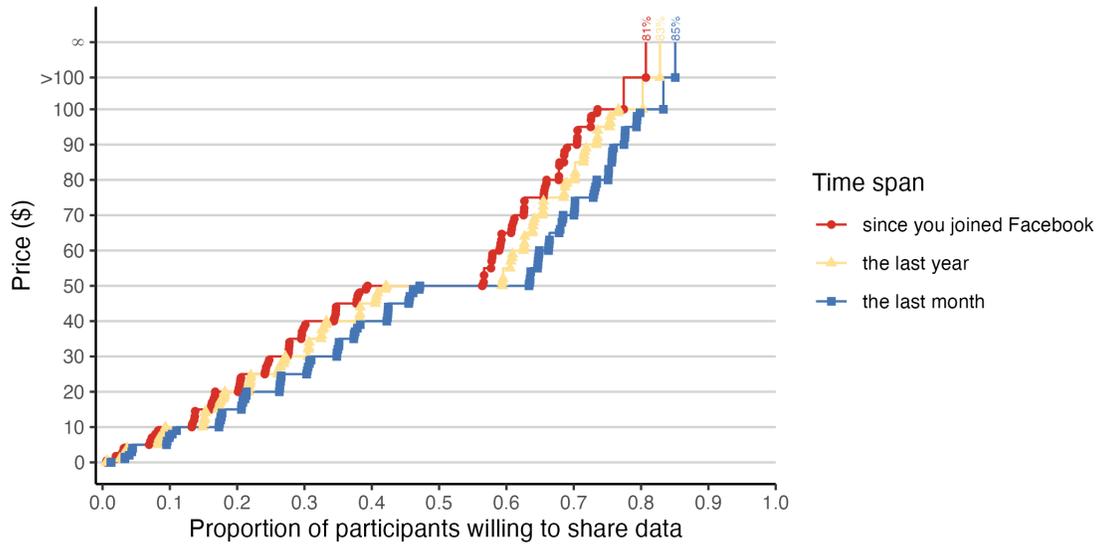

Figure B.3: Data Supply Curves by Data Duration

# C  Empirical Results Appendix

## C.1  Alternative Model Specifications (OLS and Tobit)

Table C.1: Treatment effects with OLS Specification

|                       | WTA          | log(WTA)    | WTA          | log(WTA)    |
|-----------------------|--------------|-------------|--------------|-------------|
| Intercept             | 60.438 ***   | 3.884 ***   | 48.375 ***   | 3.556 ***   |
|                       | (0.723)      | (0.022)     | (0.760)      | (0.024)     |
| Price Anchor = Low    | -16.522 ***  | -0.536 ***  | -16.522 ***  | -0.536 ***  |
|                       | (0.762)      | (0.023)     | (0.762)      | (0.023)     |
| Default = Active      | 1.763 +      | 0.042       | 1.763 +      | 0.042       |
|                       | (0.927)      | (0.029)     | (0.927)      | (0.029)     |
| Default = Opt-in      | 4.185 ***    | 0.105 ***   | 4.185 ***    | 0.105 ***   |
|                       | (0.929)      | (0.029)     | (0.929)      | (0.029)     |
| Likes                 |              |             | 6.964 ***    | 0.226 ***   |
|                       |              |             | (0.381)      | (0.012)     |
| About Me Page         |              |             | 14.833 ***   | 0.409 ***   |
|                       |              |             | (0.421)      | (0.013)     |
| Posts                 |              |             | 16.283 ***   | 0.446 ***   |
|                       |              |             | (0.429)      | (0.013)     |
| Friends and Followers |              |             | 22.236 ***   | 0.562 ***   |
|                       |              |             | (0.467)      | (0.014)     |
| Num. Obs.             | 25140        | 25140       | 25140        | 25140       |
| R2                    | 0.060        | 0.071       | 0.111        | 0.108       |

+ $p < 0.1$, * $p < 0.05$, ** $p < 0.01$, *** $p < 0.001$. The outcomes are top-coded at \$100; standard errors are clustered at the participant level.

Table C.2: Treatment Effects with Additional Controls

|  | WTA | WTA | WTA | log(WTA) | log(WTA) | log(WTA) |
|---|---|---|---|---|---|---|
| Intercept | 63.884 *** | 58.839 *** | 71.972 *** | 3.985 *** | 3.848 *** | 4.194 *** |
|  | (0.876) | (0.802) | (1.048) | (0.025) | (0.024) | (0.029) |
| Price Anchor = Low | -16.112 *** | -4.917 *** | -16.155 *** | -0.525 *** | -0.216 *** | -0.526 *** |
|  | (0.947) | (0.991) | (0.938) | (0.028) | (0.029) | (0.027) |
| Default = Active | 2.377 * | 1.930 + | 2.607 * | 0.058 + | 0.046 | 0.064 + |
|  | (1.139) | (1.029) | (1.131) | (0.034) | (0.031) | (0.033) |
| Default = Opt-in | 5.178 *** | 2.978 ** | 5.315 *** | 0.135 *** | 0.076 * | 0.139 *** |
|  | (1.147) | (1.026) | (1.135) | (0.034) | (0.031) | (0.034) |
| Practice round deviation |  | 0.695 *** |  |  | 0.019 *** |  |
|  |  | (0.021) |  |  | (0.001) |  |
| Believes data is available |  |  | -12.304 *** |  |  | -0.317 *** |
|  |  |  | (0.861) |  |  | (0.024) |
| Num. Obs. | 25140 | 25140 | 25140 | 25140 | 25140 | 25140 |

+ p < 0.1, * p < 0.05, ** p < 0.01, *** p < 0.001. The outcomes are top-coded at $100; standard errors are clustered at the participant level.

Table C.3: Average Treatment Effects and Data Valuation Across Topcodes: Tobit Regressions

|  | DV: Free-Text valuation | | | | | |
|---|---|---|---|---|---|---|
|  | 100 | 250 | 385 | 500 | 750 | 1000 |
| Intercept | 3.556 *** | 3.584 *** | 3.596 *** | 3.603 *** | 3.614 *** | 3.622 *** |
|  | (0.024) | (0.028) | (0.030) | (0.031) | (0.034) | (0.035) |
| Price Anchor = Low | -0.536 *** | -0.479 *** | -0.455 *** | -0.440 *** | -0.418 *** | -0.402 *** |
|  | (0.023) | (0.028) | (0.031) | (0.032) | (0.035) | (0.037) |
| Default = Active | 0.042 | 0.065 + | 0.075 * | 0.081 * | 0.090 * | 0.096 * |
|  | (0.029) | (0.034) | (0.037) | (0.039) | (0.042) | (0.045) |
| Default = Opt-in | 0.105 *** | 0.138 *** | 0.152 *** | 0.161 *** | 0.174 *** | 0.183 *** |
|  | (0.029) | (0.034) | (0.037) | (0.039) | (0.042) | (0.045) |
| Likes | 0.226 *** | 0.286 *** | 0.313 *** | 0.329 *** | 0.353 *** | 0.370 *** |
|  | (0.012) | (0.014) | (0.016) | (0.017) | (0.018) | (0.019) |
| About Me Page | 0.409 *** | 0.540 *** | 0.598 *** | 0.633 *** | 0.688 *** | 0.726 *** |
|  | (0.013) | (0.016) | (0.017) | (0.018) | (0.020) | (0.021) |
| Posts | 0.446 *** | 0.590 *** | 0.653 *** | 0.691 *** | 0.749 *** | 0.791 *** |
|  | (0.013) | (0.016) | (0.018) | (0.019) | (0.021) | (0.022) |
| Friends and Followers | 0.562 *** | 0.810 *** | 0.919 *** | 0.986 *** | 1.089 *** | 1.163 *** |
|  | (0.014) | (0.017) | (0.019) | (0.021) | (0.023) | (0.025) |
| Num. Obs. | 25140 | 25140 | 25140 | 25140 | 25140 | 25140 |
| R2 | 0.108 | 0.089 | 0.083 | 0.080 | 0.076 | 0.074 |

+ p < 0.1, * p < 0.05, ** p < 0.01, *** p < 0.001. The outcomes are log valuations with top-codes indicated in the first row; standard errors are clustered at the participant level.

## C.2 Causal Forest Models Tuning and Evaluation

For our heterogeneity estimation, we use the 'grf' package to implement a variety of causal forest models. We cluster the standard error at the participant level, since privacy valuations from the same individual may be correlated. In practice, clustering means that the model estimation will sample all five valuations from an individual at a time to estimate its heterogeneity. As such, the model will only pick up valuation heterogeneity that is common across the five variables, and may underestimate the general degree of heterogeneity as a result (Athey & Wager 2019). Another implication of the clustering specification is that our forest model will not recover the bunching pattern in the raw data, because bunching is rarely universal across the data valuations for a given individual.

Ex ante, we suspect that poor choices of hyperparameters may lead to poor out-of-sample fit and degraded performance of the model. To guard against this possibility, we search for the best model across three parameter dimensions: number of trees, minimal node sizes, and maximum imbalance of a split. We evaluate the model fit using the R-loss (Nie & Wager 2021), which ensures that model fit is evaluated based on the treatment effect estimation rather than its performance in predicting the outcome. We find that the choice of hyperparameters does not substantially change the qualitative heterogeneity patterns, though the choices of 'min.node.size', 'honest.fraction' and 'honest.prune.leaves' can substantially affect the R-loss. Our final model has 'min.node.size' = 3, 'num.trees = 3500', and the rest of the hyperparameters consistent with the package default value.

## C.3 Individual-Level Treatment Effect Estimates

Figure C.1: Individual-Level Treatment Effect Estimates

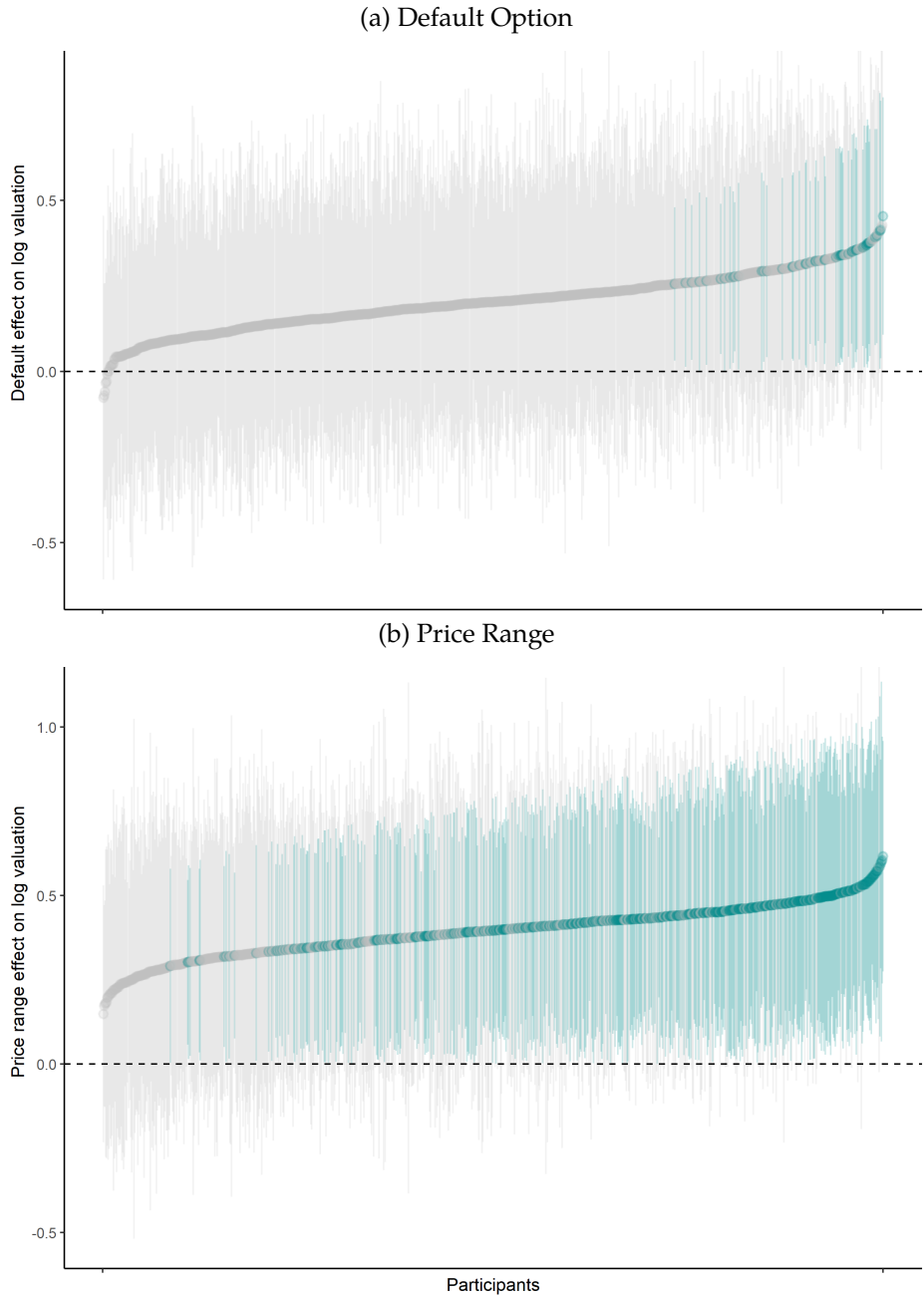

(a) Default Option

(b) Price Range

*Note:* In the figures above, dots represent HTE point estimates and vertical lines represent 95% confidence intervals; significant estimates are shaded in green. Note that The standard errors for heterogeneous treatment effects are generally larger than for the average treatment effect. In addition, clustering standard error is known to mask heterogeneity patterns substantially when combined with causal forests (Athey & Wager 2019). Therefore, it is natural for default to have a significant treatment effect in Table 2 while having insignificant effects here.

## C.4 Comparison Between Multi-Arm and Survival Forests

Since the multi-arm forest does not account for the fact that the latent censored values are greater than the censoring point, the magnitudes in its estimates are potentially biased compared to the survival forest. Figure C.2 shows that the estimated default effect distribution from the multi-arm causal forest are slightly biased upwards while the price anchor effect distribution is slightly biased downwards. Nevertheless, the magnitudes in the two models are similar, and the average treatment effects from both models are in line with estimates from our Tobit model.

Figure C.2: Heterogeneous Treatment Effect Estimates and Standard Errors: Survival Forests

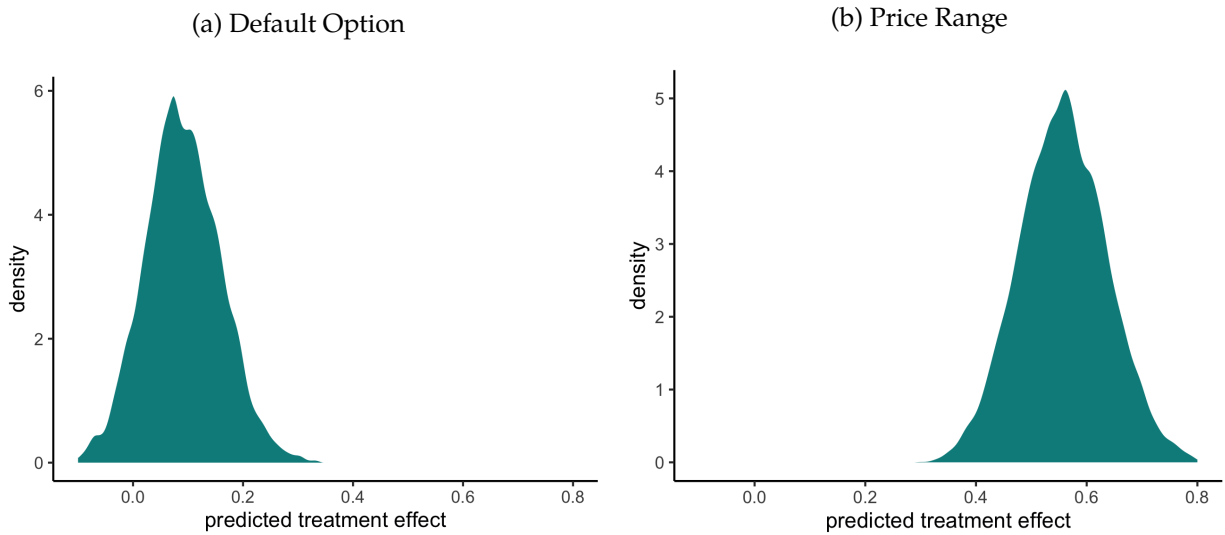

*Note:* Outcomes: log valuations truncated at $1000; standard errors are clustered at the subject ID level. ATE estimates from the causal survival forest: $ATE_{\text{default}} = 0.09$ (se = 0.03); $ATE_{\text{default}} = 0.56$ (se = 0.03).

## C.5 Treatment Effects by Sample Source

We separately compare the average log valuation and choice frame effect sizes from our two participant sources, which is shown in Figure C.3. The Facebook participants have a higher privacy valuation overall and is less influenced by frames. Although we do not know if the Facebook participants are more representative of the total population than the Prolific panel, our results suggest that studies using multiple participant sources have merit in assessing the robustness of the effects they aim to study.

Figure C.3: Average Log Valuation and Choice Frame Effect Distributions by Participant Source

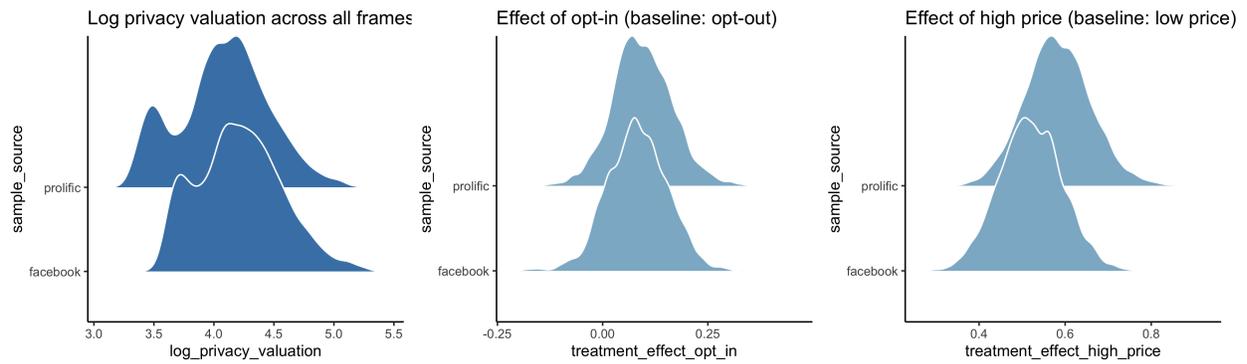

*Note:* In these figures, the treatment effects are estimated using survival forests, with the outcome as log valuation truncated at $1000.

## C.6 Elasticity Metrics Across Data Prices

Figure C.4: Elasticity Metrics Between Vol-Max and Bias-Min Frames for Each Price

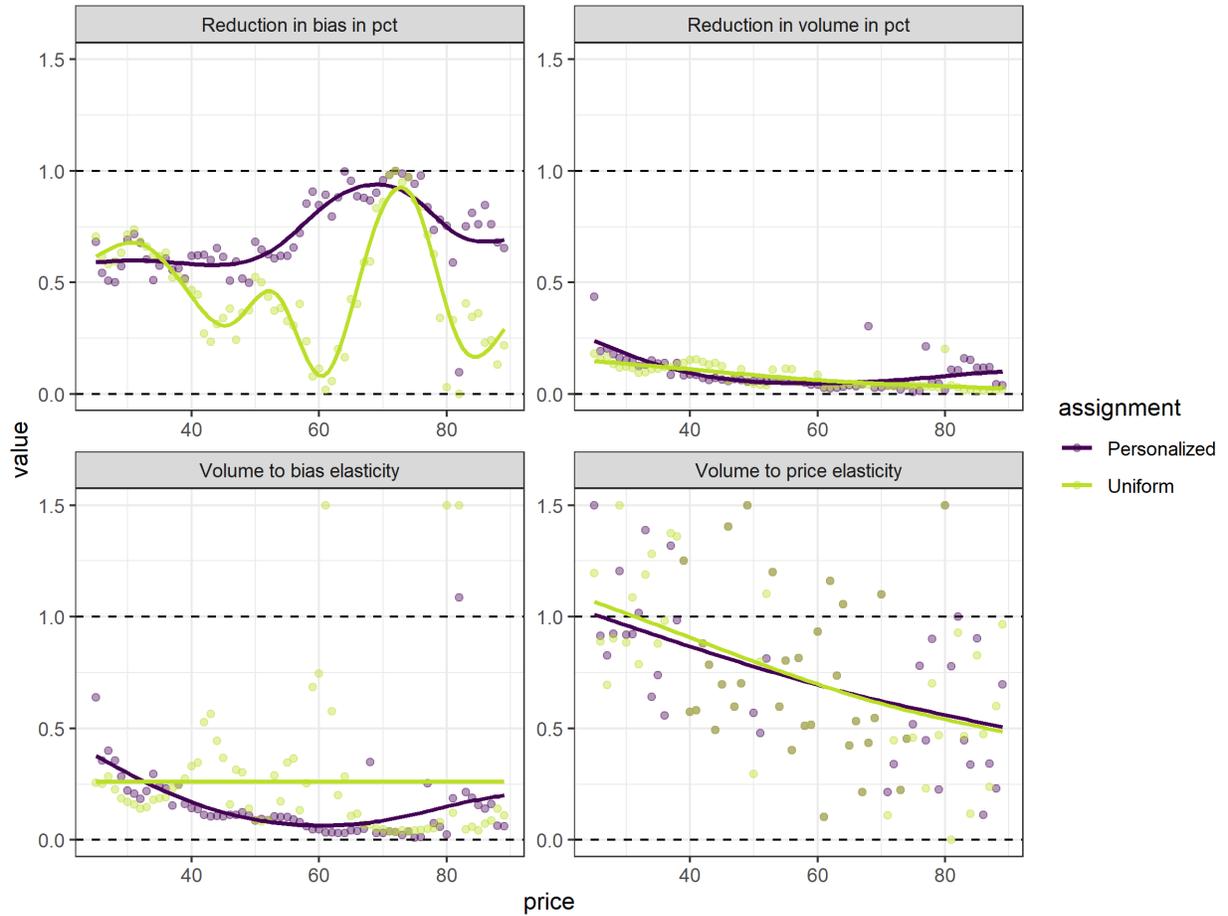

*Note:* The figure shows trade-off metrics between volume-maximizing and bias-minimizing frames at each price between $25 and $90. The green and purple colors represent the values corresponding to the uniform and personalized frame assignment, respectively.